\documentclass[symmetry,review,accept,pdftex,moreauthors]{Definitions/mdpi} 

\usepackage{graphicx}
\usepackage{amssymb}
\usepackage{amsmath}
\usepackage{tensor}
\usepackage{nicefrac}
\usepackage{amsthm}
\usepackage{mathrsfs}

\firstpage{1} 
\pubvolume{16}
\issuenum{1}
\articlenumber{13}
\pubyear{2024}
\copyrightyear{2023}
\externaleditor{Academic Editors: Olga Kodolova, Tomohiro Inagaki and Alberto Ruiz Jimeno}
\datereceived{1 December 2023} 
\daterevised{18 December 2023} 
\dateaccepted{20 December 2023} 
\datepublished{21 December 2023 } 
\hreflink{https://\linebreak doi.org/10.3390/sym16010013} 

\Title{Baryogenesis: A Symmetry Breaking in the Primordial \mbox{Universe Revisited}}

\TitleCitation{Baryogenesis: A Symmetry Breaking in the Primordial Universe Revisited}

\newcommand{\orcidJPMimoso}{{\href{https://orcid.org/0000-0002-9758-3366}{\orcidicon}}} 
\newcommand\orcidFrancisco{{\href{https://orcid.org/0000-0002-9388-8373}{\orcidicon}}}

\Author{David S. Pereira 
 $^{1}$, João Ferraz $^{1}$, Francisco S. N. Lobo $^{1,2,}$*\orcidFrancisco ~and Jos\'{e} P. Mimoso 
 $^{1,2}$\orcidJPMimoso}

\AuthorNames{David Silva Pereira, João Ferraz, Francisco S. N. Lobo and José Pedro Mimoso}

\AuthorCitation{Pereira, D.S.
; Ferraz, J.; Lobo, F.S.N.; Mimoso, J.P.}

\address{%
$^{1}$ \quad Instituto de Astrof\'{\i}sica e Ci\^{e}ncias do Espa\c{c}o, Faculdade de
Ci\^encias da Universidade de Lisboa
, \mbox{Campo Grande}, Edif\'{\i}cio C8, 1749-016 
 Lisbon, Portugal; fc54512@alunos.ciencias.ulisboa.pt (D.S.P.); jferraz@fc.ul.pt (J.F.); jpmimoso@fc.ul.pt (J.P.M.)\\
$^{2}$ \quad Departamento de F\'{i}sica, Faculdade de Ci\^{e}ncias da Universidade de Lisboa, Campo Grande, Edif\'{\i}cio C8, 1749-016 Lisbon, Portugal}

\corres{Correspondence: fslobo@fc.ul.pt}

\abstract{In this review article, we revisit the topic of baryogenesis, which is the physical process that generated the observed baryon asymmetry during the first stages of the primordial Universe. A viable theoretical explanation to understand and investigate the mechanisms underlying baryogenesis must always ensure that the Sakharov criteria are fulfilled. These essentially state the following: (i) baryon number violation; (ii) the violation of both C (charge conjugation symmetry) and CP (the composition of parity and C); (iii) and the departure from equilibrium. Throughout the years, various mechanisms have been proposed to address this issue, and here we review two of the most important, namely, electroweak baryogenesis (EWB) and Grand Unification Theories (GUTs) baryogenesis. Furthermore, we briefly explore how a change in the theory of gravity affects the EWB and GUT baryogenesis by considering Scalar--Tensor Theories (STT), where the inclusion of a scalar field mediates the gravitational interaction, in addition to the metric tensor field. We consider specific STT toy models and show that a modification of the underlying gravitational theory implies a change in the time--temperature relation of the evolving cosmological model, thus altering the conditions that govern the interplay between the rates of the interactions generating baryon asymmetry, and the expansion rate of the Universe. Therefore, the equilibrium of the former does not exactly occur as in the general relativistic standard model, and there are consequences for the baryogenesis mechanisms that have been devised. This is representative of the type of modifications of the baryogenesis processes that are to be found when considering extended theories of gravity.}


\begin{document}


\section{Introduction}

One of the greatest successes of the standard cosmological model is that it accommodates the standard model of particle physics (SM) in its framework, even though we still lack an established quantum theory of gravity~\cite{Peebles:1994xt}. In its primaeval stages, the thermal history of the Universe recreates the high-energy conditions that are required by high-energy physics~\cite{Liddle:2000cg}. Indeed, it provides a natural laboratory for the extreme range of energies under which we envisage the unification of the fundamental interactions~\cite{Dodelson:2003ft}.  
Witnessing this remarkable state of affairs, we may distinguish the accurate predictions for the cosmic microwave background (CMB), for the electron--positron annihilation occurring after neutrino decoupling, and for the big bang nucleosynthesis (BBN)~\cite{Fields:2019pfx} that predicts abundances for the light elements synthesized at the end of the first three minutes (that are in overall agreement with the primordial abundances inferred from observational data)~\cite{Mukhanov:2005sc}. 
Despite the success of the SM, there are still 
open problems, such as the late-time cosmic acceleration of the Universe~\cite{Perivolaropoulos:2021jda}, the lithium problem~\cite{Fields:2019pfx,Clara:2020efx}, and most important to this paper, the asymmetry between matter and antimatter~\cite{Linde:1990flp}. The SM predicts that all particles burst into existence following the same laws of physics, leading us to conjecture that a production of an equal amount of matter and anti-matter should have occurred, which would eventually annihilate each other, resulting in a Universe with an overall vanishing baryon number~\cite{Riotto:1999yt}.

However, cosmological observations  point to a Universe that is strongly dominated by matter over antimatter~\cite{BB, WMAP}. Without accepting some unbalanced initial conditions, there is not a simple and immediate way to solve this intriguing problem. The most logical path to take is to assume that this asymmetry was dynamically developed in the early Universe. 
With this in mind, in the modern cosmological context, baryogenesis refers to the theoretical process that occurred in the early stage of the Universe, leading to a production of an excess of matter over anti-matter. 
Over the years, various mechanisms have been proposed to address this issue. Within this landscape of baryogenesis mechanisms, categorization based on shared characteristics is feasible. This classification includes supersymmetric baryogenesis~\cite{deGouvea:1997afu,Barbier:2004ez,Ellwanger:2009dp,Chung:2003fi} (such as the well-known Affleck--Dine baryogenesis~\cite{Affleck:1984fy,Dine:1995kz}), baryogenesis via leptogenesis~\cite{Fukugita:1986hr,Luty:1992un,Davidson:2008bu,Buchmuller:2004nz}, spontaneous baryogenesis~\cite{Linde:1990flp,Albrecht:1982mp,Cohen:1991iu}, gravitational baryogenesis~\cite{Davoudiasl:2004gf}, the emerging baryogenesis through primordial black holes~\cite{Carr:2020xqk, Morrison:2018xla, Carr:2019hud,Garcia-Bellido:2019vlf, Datta:2020bht}, and even Baryogenesis from a primordial bounce induced by a repulsive potential inspired by the de Br\"oglie--Bohm quantum interpretation~\cite{Alvarenga:2001nm,Delgado:2020znu}. The most important ones that will be explored and summarized in this article are electroweak baryogenesis~\cite{Cohen:1993nk,Morrissey:2012db} and GUT baryogenesis~\cite{Kolb:1996jt,Bodeker:2020ghk}. 

In 1967, A.D. Sakharov~\cite{Sakharov} suggested for the first time that the present baryon density might not be a result from unnatural initial conditions but that it can be understood in terms of microphysical laws, which explain and dictate how an initial symmetric Universe could have evolved in a way that reached the observed asymmetry. The necessary conditions formulated by Sakharov for the generation of the asymmetry were essentially the following: (i) baryon number violation; (ii) violation of both C (charge conjugation symmetry), and CP (the composition of parity and C); (iii) and the departure from equilibrium.
The Sakharov criteria are the basic pillars to understand and investigate the mechanisms underlying baryogenesis. The first two criteria are of a quantum nature and are responsible for the ruling of the microscopic paradigm. The third one is gravitational and makes the connection between  microscopic and macroscopic scales. 

Besides these requirements, if one is assuming that the mechanisms that led to the asymmetry occurred in the early Universe, it is necessary to understand and have in mind the standard model of cosmology, which is based in Einstein’s theory of General Relativity (GR). The latter plays a crucial role in the thermal history of the Universe, which, in turn, influences the mechanisms mentioned above, in particular, electroweak baryogenesis.
In this review article, we will briefly explore how a change in the theory of gravity affects the EWB and GUT baryogenesis by considering Scalar--Tensor Theories (STT)~\cite{Avelino:2016lpj,Horndeski,Brans,Bergmann,Wagoner,Nordtvedt}. Here, the inclusion of a scalar field mediates the gravitational interaction in addition to the metric tensor field. We show that specific toy model modification of the underlying gravitational theory implies a change in the time--temperature relation of the evolving cosmological model, which consequently alters the conditions that govern the interplay between the rates of the interactions generating baryon asymmetry, and the expansion rate of the Universe. Therefore, this is representative of the type of modifications of the baryogenesis processes that are to be found when considering extended theories of gravity, or modified theories of gravity~\cite{Harko:2018ayt}.

This paper is organized in the following manner: In Section \ref{GR cosmology}, we start with an overview of the fundamentals of the standard model of cosmology. In Section \ref{thermoPU}, the thermodynamics of the primordial Universe are explored in detail. Next, in \mbox{Sections \ref{sec:SC}--\ref{sec:GUTb}}, we explore in more detail the Sakharov criteria, electroweak baryogenesis and finally GUT baryogenesis, respectively, all from the gravitational perspective.
In Section \ref{sec:STT}, the formalism of Scalar--Tensor Theories is introduced, which is used in Section \ref{sec:STTbaryogenesis} to explore how deviations from GR affect the previously studied mechanisms. Finally, in Section \ref{sec:conclusion}, we conclude by providing a comprehensive summary of the \mbox{above-mentioned mechanisms. }

\section{Standard Cosmology}\label{GR cosmology}

The standard model of cosmology is built using Einstein's GR~\cite{Mimoso:2021sic}, 
and adopts a fundamental principle denoted the Cosmological Principle, according to which the Universe is spatially homogeneous and isotropic. The latter reflects the apparent local isotropy of the large-scale observations and the absence of any reason for our position to be special.
While the adoption of GR dictates the field equations that relate the spacetime geometry to the non-gravitational fields, the Cosmological Principle endorses one to select the Friedmann--Lema\^{\i}tre--Robertson--Walker (FLRW) metric as the most adequate to characterize the gravitational field scales above $150\,$Mpc (megaparsecs). The Einstein field equations can be derived from the least action principle~\cite{Wald:1984rg,Weinberg:1972kfs}, where the fundamental action is the Einstein--Hilbert action given by
\begin{equation}\label{GR_actiom}
    S = \int d^4 x \sqrt{-g} \left[ \frac{1}{16\pi G} (R- \Lambda) + L_m \right] \ , 
\end{equation}
where \emph{g} is the determinant of the metric, $G$ is Newton’s gravitational constant, $\Lambda$ is the cosmological constant, and $\sqrt{-g} L_m$ is the Lagrangian density for the matter fields.
By varying this action with respect to the metric  degrees of freedom $g^{\mu \nu}$, 
one obtains the following field equations
\begin{equation}\label{F_Equations}
    R_{\mu \nu} - \frac{1}{2} R g_{\mu \nu} = 8 \pi G T_{\mu \nu} \ ,
\end{equation}
where $T_{\mu \nu}$ is the energy--momentum tensor of the matter fields defined as
\begin{equation}\label{T from action varition}
    T_{\mu \nu} = -\frac{2}{\sqrt{-g}} \frac{\delta (\sqrt{-g} L_m) }{\delta g^{\mu \nu}} \ .
\end{equation}


The FLRW  metric is written as 
\begin{equation}\label{FLRW}
    ds^2 = -dt^2 + a^2(t) \left[ \frac{dr^2}{1-kr^2} + r^2 (d\theta^2 + sin^2(\theta) d^2\phi) \right] \ ,
\end{equation}
where $a(t)$ is the dimensionless scale factor of the Universe, $r$, $\theta$ and $\phi$ are spatial comoving coordinates representing, respectively, a radial coordinate and the two usual spherical angular coordinates. With an appropriate rescaling of the coordinates, $k$ can be chosen to be $-1$, $0$, $+1$ if one is considering Universes with a negative, zero (flat) or positive curvature, respectively. Note that we use the metric signature convention $(-,+,+,+)$ throughout this work.  

In addition, we consider an energy--momentum tensor that is consistent with the symmetries of the FLRW metric, which takes the form~\cite{Weinberg:1972kfs}
\begin{equation}\label{Perfect_fluid}
    T_{\mu \nu} = (\rho + p) u_\mu u_\nu + p g_{\mu \nu} \ ,
\end{equation}
where $\rho$ is the energy density of the matter content of the Universe and $p$ is the pressure. Such an energy--momentum tensor corresponds to that of a perfect fluid. By plugging this energy--momentum tensor into the Einstein field Equation \eqref{F_Equations}, and taking into account the FLRW metric, one obtains the following dynamical equations:
\begin{equation}\label{Friedmann}
    H^2(t) + \frac{k}{a^2(t)} = \frac{8\pi G}{3}\rho + \frac{\Lambda}{3} \,,
\end{equation}
\begin{equation}\label{aceleration}
    \dot{H}(t) + H^2 (t) = - \frac{4\pi G}{3}(\rho + 3p) + \frac{\Lambda}{3} \,,
\end{equation}
respectively,
where $H \equiv \dot{a}(t)/a(t)$ is the expansion rate of the Universe, where the overdot denotes a derivative with respect to cosmic time $t$. Equation \eqref{Friedmann} is denoted as the Friedmann equation and is obtained from the $00$ component of the Einstein's field equations. \mbox{Equation \eqref{aceleration}} is the Raychaudhuri equation, which is an equation for the acceleration and 
is found by subtracting the Friedmann equation from the $i-i$ components of the Einstein \mbox{field equations.} 

The Friedmann equation can be recast in a 
particularly suitable form through the definition of two dimensionless 
parameters: the critical density, a parameter that can be found by setting $\Lambda = k = 0$ in the aforementioned equation, and it is therefore defined as 
\begin{equation}\label{critical density}
    \rho_c = \frac{3H^2(t)}{8 \pi G} \ ,
\end{equation}
and the dimensionless density parameter, which is given by the ratio of the energy density $\rho$ to the critical density ($\rho_{c}$) at a given time
\begin{equation}\label{density parameter}
    \Omega(t) \equiv \frac{\rho(t)}{\rho_c(t)} \ .
\end{equation}

With these definitions, one can rewrite the Friedmann \eqref{Friedmann} in the following form:
\begin{equation}\label{Omega}
    \frac{k}{a^2H^2(t)} = \Omega(t) - 1 \ .
\end{equation}

This establishes a correspondence between the sign of $k$ and the sign of $\Omega - 1$, i.e., considering $k = 0$ (flat model) leads to $\Omega = 1$, for $k = 1$ (closed model) leads to $\Omega > 1$ and if $k=-1$ (open model) leads to $\Omega<1$. Besides this, the Einstein field equations entails the following conservation equation:
\begin{equation}\label{Conservação}
    \dot{\rho} = -3 H(t)(\rho + p) \ ,
\end{equation}
which reflects the conservation of energy and is intrinsically contained in the structure of the gravitational field equations. This conservation comes from the fact that the Einstein tensor $G_{\mu \nu}$ satisfies the contracted Bianchi identities, which in turn leads to the conservation of the energy--momentum tensor, i.e., $\nabla_{\mu}T^{\mu \nu} = 0$. The component $\nu = 0$ of this conservation results explicitly in Equation \eqref{Conservação}. To close the system of equations governing the evolution of the scale factor, it is necessary to introduce an adequate barotropic equation of state, i.e., $p=w \rho$. For all the components, the most simple equation of state is provided by 
\begin{equation}\label{baritropic PFluid}
    \sum_i p_i = \sum_i (\gamma_i-1) \rho_i \ ,
\end{equation}
where $\gamma_i$ is a constant parameter, and $\gamma_i-1$ is essentially the speed of sound $v^2_s$ in the fluid $i$ that characterizes the $i$-th component of the matter content of the Universe. This equation of state allows to integrate the equation for the conservation of energy, \mbox{Equation \eqref{Conservação}}, leading to
\begin{equation}\label{densidade}
    \rho (t) = \sum_{i} \rho_{i,0} \, \left(\frac{a(t)}{a_0}\right)^{-3\gamma_i} = \sum_i \rho_{i,0} (1+z)^{-3\gamma_i} \ ,
\end{equation}
where $\rho(t) \equiv \sum_i \rho_i(t)$ represents the total energy density of the Universe at a given time $t$, and 
$\rho_{i,0} \equiv \rho_i(t_0)$  and $a_0 \equiv a(t_0) $ are constants of integration which, without loss of generality, can be respectively established by the values of each component $i$ of the matter content of the Universe and the value of the scale factor at the present day $t=t_0$. Therefore, using Equation \eqref{densidade} and taking into account that the present day value of the critical density parameter is given by $\rho_{c,0} = 3 H^2_0/(8 \pi G)$, one can write the density parameter in terms of the redshift $z$ as follows:
\begin{equation}\label{density_parameter_redshift}
    \Omega(z) = \frac{H^2_0}{H^2(z)} \left[\sum_i \Omega_{i,0} (1+z)^{3\gamma} + \Omega_{\Lambda} \right] \ ,
\end{equation}
where $\Omega_{\Lambda} \equiv \Lambda/(3H^2_0)$ and $\Omega_{i,0} \equiv \rho_{i,0}/\rho_{c,0}$ are, respectively, the contributions from the cosmological constant and from the $i$-th fluid component for the present day \mbox{density parameter.  }


To inspect the dynamical behavior of the scale factor during the evolution of the Universe, the general approach consists in dividing the history of the Universe into several time domains or cosmological epochs, during which one single fluid characterized by a constant parameter $\gamma_i$ dominates. The usual cases of interest include the radiation fluid ($i = r$), where $p = \rho$/3 ($\gamma_r = 4/3$); incoherent matter ($i = m$), where $p = 0$ ($\gamma_m  = 1$); and the vacuum energy ($i = \Lambda$ ), characterized by $p = -\rho$ ($\gamma_\Lambda = 0$), which is therefore physically equivalent to a cosmological constant. Therefore, from the equation for the conservation of energy, Equation \eqref{densidade}, the density profile for each of these epochs becomes simply characterized by $\rho(t) \simeq \rho_i(t) \propto a(t)^{-3\gamma_i}$ and from the substitution of this profile in the Friedmann equation, Equation \eqref{Friedmann}, it becomes clear that for $\gamma_i > 2/3$, the curvature term $k^2/a^2$ only dominates the expansion at a rather late time and only if a cosmological constant does not dominate sooner. Thus, regarding the alluded cases of interest, it is a good option, at least in what concerns
the first moments of the Universe, to consider only the flat model case, which then allows the Friedmann Equation \eqref{Friedmann} to be integrated, yielding cosmological solutions characterized by 
\begin{equation}\label{S1}
    a(t) \propto t^{\frac{2}{3\gamma_i}} ; \ H(t) = \frac{2}{3\gamma_i} \frac{1}{t} ; \ \ \  if \  \gamma_i \neq 0 \ , 
\end{equation}
\begin{equation}\label{Sitter}
    a(t) \propto e^{\sqrt{\frac{\Lambda}{3}}t} ; \ H(t) = \sqrt{\frac{\Lambda}{3}} ;  \ \ \   if \ \gamma_i = 0 \ . 
\end{equation}

\section{Thermodynamics in the Primordial Universe}\label{thermoPU}

The previous treatment of the Universe allowed us to derive the basic equations that govern the expansion of the Universe. However, to fill in the details of the history of the Universe and to attain a better comprehension of the mechanisms behind the breaking of the matter and anti-matter symmetry, it is necessary to follow the thermal evolution of its leading constituents, using thermodynamics, both in and out of equilibrium, thus having a notion of how they behave and interact. Most of the early history of the Universe can, in good approximation, be well described in terms of equilibrium thermodynamics~\cite{Bernstein}. With this in mind, in this section, all the fundamentals to make this type of description will be explored below.

\subsection{Thermal History of the Universe}

The thermal history of the Universe is fundamental to understand when and how the baryogenesis mechanisms occur.
~The timeline of thermodynamic events in the Universe not only allows us to frame these mechanisms for breaking baryonic symmetry but also to understand more complex dynamics related to gravitation itself.  With this being said, the thermal history of the Universe presents the following scenario: 

\begin{enumerate}

    \item \emph{Quantum Gravity?} 
 ($T>T_{Pl}\sim 10^{19}$ GeV)---GR is rendered invalid at this scale so that quantum corrections are necessary. In inflationary scenarios, this is usually called pre-inflationary cosmology.

   \item \emph{Inflation}---This is an epoch of the accelerated expansion of the Universe, probably exponential. In such a case, the Universe can be described by the de Sitter solution, which is characterized by Equation \eqref{Sitter}. Adiabaticity fails, and equilibrium thermodynamics is not valid in such a period of the history of the Universe. 

   \item \emph{End of inflation and particle production}---In this period, dark expanding ``emptiness'' filled by a scalar (or some other) field, possible inflaton, creates radiation and other elementary particles. This is the period of reheating. 

    \item \emph{Start of the radiation-dominated Universe}---The Universe becomes well described by equilibrium thermodynamics, being adiabatically cooled down. If current ideas are correct, during this epoch the Universe underwent several phase transitions. During the stages at which phase transitions occur, adiabaticity could be broken. 

    \item \emph{Grand unification phase transition} ($T\sim 10^{15}$ GeV--$10^{17}$ GeV)---The main goal of the Grand Unification Theories (GUTs) is to attempt to unify the electromagnetic, electroweak and strong interaction into a simple gauge group that should be a valid symmetry at the highest energies. As energy is lowered, the theory undergoes a hierarchy of spontaneous symmetry breakings (SSBs) into successive subgroups. From the cosmological point of view, these SSBs should have been reflected into phase transitions occurring during the evolution of the Universe, with the possible formation of \mbox{topological defects. }

    \item \emph{EW phase transition} ($T\sim 100$ GeV)---Some of the gauge bosons and other particles acquire mass via the Higgs mechanism. 

    \item \emph{QCD phase transition} ($T\sim 150$ MeV)---Quarks lose their so-called asymptotic freedom,
which they have at high energies, and there are no more free quarks and gluons, rather the quark--gluon plasma becomes a hadron gas. Combinations of three quarks (baryons) and quark--antiquark pairs (mesons) are formed.

    \item \emph{Neutrino decoupling} ($T\sim 1$ MeV)---Previous to this point in the history of the Universe, neutrinos were kept in thermal equilibrium via the weak interactions of the type $\bar{\nu}\nu\leftrightarrow e^{+}e^{-}$, where $\nu$ and $\bar{\nu}$ represent, respectively, generic neutrinos and antineutrinos, and $e^{+}$ and $e^{-}$ represent positrons and electrons, respectively.

    \item \emph{Electron--positron annihilation} ($T\sim 0.5$ MeV)---Shortly after the neutrino decoupling, the temperature drops below the mass of the electron/positron, which thereby become non-relativistic.

    \item \emph{Big Bang Nucleosynthesis} ($T\simeq 10$--0.1 MeV)---At about $1$ MeV, the ratio between the neutron number and the proton number (usually termed neutron-to-proton) ratio freezes out. Shortly thereafter ($T\sim 0.1$ MeV), the synthesis of light elements begins.

    \item \emph{Matter--radiation equality}---After nucleosynthesis, the Universe reaches a point after which matter begins to dominate over the radiation, relativistic particles that, within FLRW models, have an energy density proportional to $a^{-4}$ or to $T^4$ since $T \propto a^{-1}$ (as the scale factor grows, the temperature decreases). According to the stages of the Universe previously defined, such a point marks the entrance of the Universe in its adolescence, and it is usually called the matter--radiation equality.

    \item \emph{Photon decoupling and recombination} ($T\sim 0.2$ MeV--$0.3$ MeV)---The interactions between photons and electrons are rapid relative to the expansion rate of the Universe, so radiation (photons) and matter (electrons, protons and nuclei) are kept in good thermal contact, thus remaining in equilibrium. However, eventually, the Universe reaches a point were the thermal contact between these two components is no longer maintained and radiation decouples from matter.

    \item \emph{The Dark Age}---Here, stars are not yet formed, and with the expansion of redshift cosmic microwave background photons towards the infrared, they become invisible heat radiation. 

    \item \emph{Formation of the first stars and reionization}---In the dark, the seeds of structure formation were already settled, and as masses were gathering together, one by one the first stars lit up.

    \item \emph{Present time $t_0$}. 
    
\end{enumerate}

\subsection{Kinetic Equilibrium: Distribution Functions for Bosons and for Fermions}

In thermodynamics, after a sufficiently long time, an isolated system usually attains a state of thermal equilibrium with all its macroscopic properties, such as particle distribution, assuming their most probable values. In these cases, not only the density number $n$ but also all the other relevant thermodynamic quantities, such as the energy density $\rho$, pressure $p$, and entropy density $s$, are given as integrals over the distribution function $f(x^a,p^a)$. Using the FLRW metric, the phase space density must be isotropic and homogeneous in agreement with the background metric so that $f(x^a,p^a) = f(|\vec{\textbf{p}}|,t) = f(E, t)$, where  $E = \sqrt{\vec{\textbf{p}}^2 + m^2}$ is the energy. Therefore, leaving its time dependence implicit, which will be manifest via its temperature dependence, the phase space distribution function for a given particle species $i$ in kinetic equilibrium in the ideal gas approximation reduces to the Fermi--Dirac (FD) or Bose--Einstein (BE) distributions 
\begin{equation}\label{D_function}
    f_i(\vec{p}) = \left[ exp(\frac{E_i(\vec{p}) - \mu_i}{T})\pm 1 \right]^{-1} \ ,
\end{equation}
where $E_i$ is the energy of the particle species $i$, $m_i$ is its corresponding rest mass, $\mu_i$ is the particle species chemical potential, and the $(+)$ sign corresponds to fermions, that is, particles which obey FD statistics and the $(-)$ sign corresponds to bosons, that is, particles that satisfy the BE statistics.
If one drops the term $\pm 1 $, then one recovers the classical and  distinguishable particles approximation of Maxwell--Boltzmann (MB) statistics. Although this does not result in an exact distribution function, it is often of interest because in the absence of a degenerate Fermi species, that is, $\mu_i \gtrsim T_i)$ or a Bose condensate, the use of MB statistics introduces only a small quantitative change relative to the exact statistics~\cite{Kolb:1990vq}, corresponding to an error that is often less than 10\%~\cite{J.C. Mather}. One of the advantages of this approximation resides in the cases where considering non-relativistic species of particles, i.e., $m_i \gg T_i$, with $m_i \gg T + \mu_i$ leads to the MB statistics becoming exact.

\subsection{Chemical Equilibrium}

The chemical potential provides generically the infinitesimal change in energy by adding a new particle of a particular type and, although usually unknown, the fundamental relation of thermodynamics, which for the general case of a variable number of particles assumes the form
\begin{equation}\label{Entropia}
    dE =  T dS - p dV + \sum_i \mu_i dN_i \ ,
\end{equation}
where S is the entropy, $V$ is the volume, and $N_i$
is the particle number of the species $i$. This allows to establish that, for any interaction frequently taking place in an equilibrium gas, one has $\sum_i \mu_i dN_i = 0$. That is so because otherwise, the number of particles of different species would change to lower the free energy given by $F = E - T S $, where $E$ is the internal energy, which at constant $T$ and $V$ obeys the differential relation 
\begin{equation}
    dF = \sum_i \mu_i dN_i \ , 
\end{equation}
and thus it would not be in equilibrium; therefore, if complete equilibrium holds, i.e., kinetical plus chemical equilibrium, then not only are the particle distribution functions of the equilibrium gas given by their usual thermal forms of Equation \eqref{D_function} but also the chemical potential of one species is related to the chemical potentials of other particle species with which it interacts. More precisely, whenever the chemical equilibrium is maintained, $\mu_i$ is additively conserved in all reactions so that if, for example, equilibrium with respect to the reaction $a_1 + a_2 + ... \leftrightarrow b_1 + b_2 + ...$ is maintained, then the chemical potentials of the particles participating in the reaction are related by
\begin{equation} \label{ Chemical equality}
    \sum_k \mu_{ak} = \sum_{k} \mu_{b_k} \ .
\end{equation}

Yet, relative to radiation, photons can be emitted or absorbed in an arbitrary reaction in any number, and for any charged particle, there is an inelastic scattering reaction $i + i \to i + i + \gamma$, where $i$ represents a particle species. The existence of these reactions implies that whenever chemical equilibrium is maintained, $\mu_i + \mu_i = \mu_i + \mu_i + \mu_\gamma$ and so, in complete equilibrium, the photon has zero chemical potential. With this result in mind, when considering the reactions of the annihilation of particles and antiparticles $i + \overline{i} \leftrightarrow 2\gamma, 3\gamma, ...$ in equilibrium, these reactions imply that the chemical potentials of particles and antiparticles are related by
\begin{equation}
    \mu_{i} + \mu_{\overline{i}} = 0 \ ,
\end{equation}
allowing to conclude that the chemical potentials of particles and antiparticles are of equal magnitude, albeit of opposite signs. Also, in the particular cases where particle species are self-conjugated ($i = \overline{i}$) or there is a symmetry between particles and antiparticles ($n_i = n_{\overline{i}}$), chemical equilibrium concerning the annihilations implies that $\mu_{i} = \mu_{\overline{i}}$ = 0. Thus, it becomes clear that under these situations, the distribution functions are determined by one parameter only, namely, the temperature. In other words, the distribution function goes from $f(x^a,p^a)$ to $f(T)$. 

It is useful to introduce chemical potentials associated with conserved quantum numbers $Q(a)$ such that the chemical potential, for a given particle species $i$, is given by
\begin{equation}\label{Chemical Quantum}
    \mu_i = \sum_k \mu_k Q_i^{(k)} \ ,
\end{equation}
where $Q_i^{(k)}$ represents the quantum numbers carried by particle $i$, which should be independent of each other and form a complete set of conserved quantum numbers~\cite{A.D. Dolgov}. So, if quantum numbers are conserved in all reactions, then the relation present in Equation \eqref{ Chemical equality} holds automatically, and all chemical potentials can be expressed in terms of the chemical potentials of these conserved quantities. Therefore, there are as many chemical potentials as there are independent conserved quantum numbers, and if charge densities
$n_{Q^{(k)}}$ are known, then the chemical potentials to the charges $Q^{(k)}$ can be found. Indeed, the number density of each particle species $n_i$ can, in this case, be given as a function of $\mu_i$, so the system of equations
\begin{equation}
    \sum_i Q_{i}^{(k)} n_i = n_{Q^{(k)}} \ ,
\end{equation}
together with Equation \eqref{Chemical Quantum} determines all $\mu_i$ in terms of $n_{Q_i}$~\cite{A.D. Dolgov}.

\subsection{Interactions in an Expanding Universe}

Relatively to the coupling or decoupling of a particle species, a useful criterion is provided by the comparison of the interaction rate $\Gamma_{int}$ for the processes in which the particle species participate with the expansion rate of the Universe. The two relations of interest between these two quantities are the following:
\begin{equation}
\label{Interactions vs H}
    \Gamma_{int} > H \ , \ coupled
\end{equation}\vspace{-6pt}
\begin{equation}
    \Gamma_{int} < H \ , \ decoupled
\end{equation}

In an expanding Universe, thermal equilibrium can be hardly maintained because one must take into account that the temperature and the chemical potential must change so that energy  and particle number conservation are satisfied. So, in principle, an expanding Universe is not in equilibrium~\cite{Bernstein}. Despite this, the expansion of the Universe is, in general, so slow that the particle soup usually has time to settle close to local equilibrium, and since the Universe is homogeneous, local values of thermodynamic quantities are also global values. Thus, if particle interaction rates occur rapidly compared to the expansion rate of the Universe, then equilibrium particle distributions will be established on the expansion time scale. So, in the case of a given particle species, we can qualitatively assume that this species will pass through a succession of equilibrium-like states whenever the condition (\ref{Interactions vs H}) is satisfied. 
To compliment this, one can also consider the condition, $p^0 < m_{Pl}$, where $m_{Pl}$ is the Plank mass~\cite{Bernstein}. This consists, in fact, of a pretentious rule of thumb, as the thermal history of the Universe can be seen as a competition between the particle interaction rates and the expansion of the Universe.
From Equation \eqref{S1}, it can be seen that for flat models with $\gamma_i \neq 0$, the age of the Universe is proportional to $H^{-1}$ so that the condition $\Gamma_{int} > H$ is roughly equivalent to requiring that, on average, at least one interaction involving the particle species under consideration has occurred over the lifetime of the Universe~\cite{Kolb:1990vq}.

\subsection{Entropy in the Expanding Universe and Baryon Number}

In the expanding Universe, entropy plays a central role, as it provides an essential criterion to follow the scale factor in terms of temperature. This state variable enters the fundamental relations of thermodynamics which, in the general case of a variable number of particles, takes the form present in Equation \eqref{Entropia}. Because the energy and the number of particles are extensive quantities (proportional to the volume of the system) and the temperature and pressure are intensive quantities and hence are independent of the volume, entropy must be an extensive quantity. Therefore, it is possible to rewrite the fundamental relation of thermodynamics in terms of the energy, number and entropy densities in a cosmological volume \emph{V} so that it may be rewritten as
\begin{equation}
    (Ts - \rho - p + \mu n)dV + (Tds - d\rho + \mu dn) V = 0 \ ,
\end{equation}
where $\rho = E/V$ is the energy density, $n=N/V$ is the number volume, and $s=S/V$ is the entropy density. 
By the second law of thermodynamics, the entropy for any closed system can only increase, and it stays constant for equilibrium or adiabatic evolution; in other words, a slow evolution during which the systems always remains in thermal equilibrium. 
Thus, considering an expanding Universe, applying this fundamental relation of thermodynamics to a comoving volume defined as V $\equiv a^3$, one finds that in thermal equilibrium, the variation of the entropy in a comoving volume $S=sV$ is given by
\begin{eqnarray}
    \frac{dS}{dt} &=& \frac{d}{dt}\,\Big[\frac{a^3}{T}(\rho+p-\mu n)\Big]
    \nonumber \\
    &=&\frac{a^3}{T}\,\Big[3H(\rho+p)+\frac{d\rho}{dt}+\frac{dp}{dt}-s\frac{dT}{dt}-n\frac{d \mu}{dt}-\mu\Big(\frac{dn}{dt}-3H n\Big)\Big]  \ .
\end{eqnarray}

Recognizing that for vanishing chemical potentials $\dot{p}=s\dot{T}$, it follows from the conservation of energy Equation \eqref{Conservação} that for vanishing chemical potentials, one arrives at
\begin{equation} \label{conservation of the entropy a}
\frac{dS}{dt}=0 \ . 
\end{equation}

With this result, it becomes evident that the conservation of energy possesses a simple interpretation when analyzed from the entropy perspective. This result is remarkably general: on the one hand, it is also true for any particle species with distribution function given by $f=f^{\textrm{eq}}(E/T)$ and arbitrary $T(t)$ that satisfies the equation for the conservation of energy~\cite{A.D. Dolgov, Dmitry S. Gorbunov}. Thus, it also applies separately to any decoupled stable particle species if they are each separately in a thermal distribution with their temperature. On the other hand, returning to the fundamental relation of thermodynamics and recalling that it is convenient to introduce chemical potentials to conserved quantum numbers rather than to individual particle species so that $dN_i$ should be understood as differentials of these conserved quantum numbers, it is found that $dN_i=0$ whenever such quantum numbers are indeed conserved, and thus in this situation, the previous results remain valid even for non-vanishing chemical potentials~\cite{5C}.

As a consequence, it can be seen that the conservation of entropy per comoving volume can also be used to introduce quantitative time-independent characteristics of asymmetries of conserved quantum numbers~\cite{A.D. Dolgov,5C}. Thus, for practical purposes, there is an asymmetry between matter and antimatter and, if there are no processes that violate the baryon number, this asymmetry must be conserved in a comoving volume so that
\begin{equation}
    (n_b-n_{\bar{b}})a^3=cte \ ,
\end{equation}
where $n_b$ and $n_{\bar{b}}$ are the number densities of baryons and anti-baryons, respectively, and it is possible to define the following ratio:
\begin{equation}
    B \equiv \frac{n_B}{s} \ ,
\end{equation}
where $n_B\equiv n_b-n_{\bar{b}}$, is a time-independent characteristic of the baryon asymmetry provided that the expansion remains isentropic.
This is usually true with very good accuracy during most of the Universe's history. After the photon decoupling, the number of non-interacting photons in a comoving volume also remains constant $T_{\gamma}\propto a^{-1}(t)$, $n_{\gamma}\sim a^{-3}(t)$ and so, after this point, as long as there is no more violation of the baryonic charge, the ratio $N_{B}/N_{\gamma}=n_{B}/n_{\gamma}$ does not change either. 
However, at higher temperatures, there is an annihilation of the massive particles whenever the temperature drops below their masses, and these annihilations heat the primordial plasma and increase the photon number density. Because of that, it is more convenient to introduce the quantity $B$, which in a thermal equilibrium state, remains effectively constant during the course of expansion. Yet, in the standard model of particles (SMP), without new long-lived particles, there is no longer a transfer of entropy to the photons after the $e^+e^-$ annihilation and so, $n_{\gamma}a^3$ also remains constant. Thus, at lower temperatures, namely $T\lesssim 1$ MeV, one traditionally uses the \emph{baryon-to-photon} ratio
\begin{equation}
\eta\equiv \frac{n_B}{n_{\gamma}} \ ,
\end{equation}
which for this temperature range also remains constant and roughly equal to $10 B$ so that the two quantities practically coincide with each other. At higher temperatures though, they may differ by one to two orders of magnitude due to the contribution of heavier particles to the entropy density so that the baryon-to-photon ratio is diluted by the same amount. Other sources for the dilution of the baryon-to-photon ratio are possible phase transitions in the early Universe of the first and even of the second order, or out-of-equilibrium decay of unstable particles. These later possibilities could also considerably diminish the original baryon asymmetry characterized by $B$.

\subsection{Equilibrium in an expanding Universe}
\label{Equilibrium in the Expanding Universe}

At high temperatures, usually associated with the early Universe, the chemical potentials are expected to be very small so that as a first approximation, they can be neglected. Under this assumption, the computations become greatly simplified since the thermal distribution functions and all thermodynamic quantities become dependent only on the temperature. Therefore, whenever equilibrium is maintained between the main components of the matter content of the Universe, it is a convenient and a good approximation to include only relativistic particles in the total thermodynamic quantities so that the expansion rate of the Universe can be given by
\begin{equation}\label{radiation thermal expansion rate}
    H= 1.66 g_{\ast}^{1/2}(T)\frac{T^2}{m_{Pl}}\ ,
\end{equation} 
where $g_{\ast}(T)$ counts the effectively massless degrees of freedom and is given by
\begin{equation}
    g_{\ast}(T)=\sum_{i=\textrm{bosons}}g_i\,\left(\frac{T_i}{T}\right)^4+\frac{7}{8}\sum_{i=\textrm{fermions}}g_i\,\left(\frac{T_i}{T}\right)^4\ .
\end{equation}

It 
 is also important to note that $g_{\ast}(T)$ always depends on the temperature since as it varies, particle species may become relativistic or non-relativistic, adding or dropping the sum, respectively. The value of $g_{\ast}(T)$ at any given temperature depends also on the particle physics model, and $\sqrt{g_{\ast}}$ varies between $2$ and $20$. For instance, in the standard $\textrm{SU}(3)\times \textrm{SU}(2)\times \textrm{U}(1)$ model, it is possible to specify $g_{\ast}(T)$ up to temperatures of $\textrm{O}(100)$ GeV, while at higher temperatures, $g_{\ast}(T)$ will be model dependent. In the minimal $\textrm{SU}(5)$ for $T\gtrsim 10^{15}$ GeV, $g_{\ast}(T)=160.75$.

One can also consider that if the equilibrium situation is indeed maintained, it is expected that the Universe is at its earliest times a very high-temperature system mostly filled with relativistic particle species ($T\gg m_i$) and consequently, all interactions should be mediated by massless gauge bosons with corresponding cross sections given by $\sigma_k\sim \alpha^n T^{-2}$, where $\alpha$, usually within the range $0.01<\alpha<0.1$, is the generic value of the coupling constant relevant for the process under consideration and $n=1,~2$ stands for decays and two-body reactions, respectively. Thus, in this regime, the scattering cross sections for the different types of processes are all approximately given by the same expression. This also holds for the interaction rates per particle, which are given by
\begin{equation}
    \Gamma\equiv n_T\sigma \vert v\vert\ ,
\end{equation}
where $v$ is the relative velocity of the particles  and $n_T$ is the number density of the target particles.

\section{Sakharov Criteria}\label{sec:SC}

In what concerns baryogenesis, it was first suggested by A.D. Sakharov, in a seminal paper from 1967, that the baryon density might not represent an unnatural initial condition but might be understandable in terms of microphysical laws, which explain and dictate how an initial symmetric Universe could have dynamically evolved the observed asymmetry. As mentioned in the Introduction, Sakharov formalized the following conditions necessary to generate asymmetry:
\begin{enumerate}
\item Baryon number violation;
\item Violation of C (charge conjugation symmetry) and CP (the composition of parity \mbox{and C);}
\item Departure from the equilibrium.
\end{enumerate}     

\subsection{Baryon Number Violation}

The first condition is clear since starting from a symmetric Universe, baryon number violation must take place for the Universe to evolve into a state in which $\eta$ is non-zero. As a process that ends up generating $\Delta B > 0$, an example of this process can be expressed as $X \to Y + b$, leading to $\Delta B = 1$.

\subsection{Violation of C and CP}

The second Sakharov criterion is necessary because if C and CP were satisfied, in the absence of a preference for matter or antimatter, baryon number-violating interactions would produce baryon and antibaryon excess at the same rate, thereby maintaining a zero net baryon number. This is the same as saying that the thermal average operator B, which is an odd operator under both C and CP transformations, is zero unless these symmetries are violated. For the last condition, one can calculate the equilibrium average of $B$ as the following~\cite{Riotto:1999yt}:
\begin{eqnarray}
\left\langle B\right\rangle_T=Tr[e^{-\beta H}B]=Tr[(CPT)(CPT)^{-1}e^{-\beta H}B]\nonumber\\
=Tr[e^{-\beta H}(CPT)(CPT)^{-1}]=-Tr[e^{-\beta H}B]\,,
\end{eqnarray}
so, in equilibrium, $\left\langle B\right\rangle_T=0$, and there is no generation of a net Baryon number. 

\subsection{Departure from Equilibrium}

From all the conditions, what is important for the main goals of this review is the departure from equilibrium. This departure can be directly or indirectly provided by the expansion of the Universe, linking the macroscopic physics, from which the dynamics of the Universe are derived, with the microphysics that explains the unnatural asymmetry between matter and anti-matter and consequently for the observed baryon number as a consequence of an interplay between the laws of the macrophysics and microphysics. In thermal equilibrium, we also have that the process  $X \to Y + b$ is compensated by its counterpart  $Y + b \to X$, leading to a total baryon number variation of zero. 

\section{Eletroweak Baryogenesis}

Electroweak baryogenesis (EWB) is one of the most well-known and promising ways to account for the observed baryon asymmetry of the Universe. As its name suggests, this theory refers to the mechanism that produces an asymmetry in the number of baryons during the electroweak phase transition of the primordial Universe. 
In the SM of particles, it was thought that the first two Sakharov criteria were not satisfied. So, a baryon number violation could not be accomplished within this theory. This reasoning comes because, at the classical level, baryon number is indeed conserved in the SM, that is, $\partial_{a} j_B^a=\partial_{a} j_L^a=0$, where $j_B^{a}=\bar{q}\gamma^a q$, $j_L^{a}=\bar{l}\gamma^a l$ are respectively the baryonic and leptonic vector currents, which means that the baryonic and leptonic charges given respectively by $B=\int\, d^{3}xj^0_B$, $L=\int\, d^{3}xj^0_L$, where $j^0_{B,L}=\rho_{B,L}$, can be easily seen to be time independent using current conservation, and thus baryon number violation cannot be easily checked out. However, this previous conservation does not hold when quantum corrections are considered. As long as quantum fluctuations around the local value $W_{\bar{\mu}}$ are small, then $\Delta B = 0$, but there are large fields that can violate the baryon number. In $1976$, 't Hooft~\cite{Hoof} showed that these special field configurations, known as instantons, can tunnel through the barrier from a state $\vert \varphi,~B\rangle$ to a state $\vert \varphi,~B+3\rangle$ with probability per unit volume $\left \langle \varphi,~B+3\vert\varphi,~B\right \rangle\propto e^{-4\pi/\alpha_W}\sim 10^{-164} $, where $\alpha_W=g^2_W/4\pi$ is the weak coupling constant. This result is valid at zero temperature and explains why baryon number violation has never been observed in experiments. At finite temperatures, this picture changes, and field configurations exist that could ``jump'' over the barrier from one vacuum to another. These configurations are saddle point solutions of the classical field equations which sit on the top of the barrier and are known as \emph{sphalerons}, from the Greek ``ready to fall''. The rate for such a process is given by the calculation of the fluctuations of the sphaleron around the saddle point. The rate per unit volume and unit time was calculated by Arnold and MacLerran~\cite{Arnold:1987mh} and is given by
\begin{equation}
\Gamma_{sph}/V= a(E_{sph})^3 m_W(T) \exp{\frac{-E_{sph}}{T}} \ ,
\end{equation}
where $m_W(T)=g_W v/2$ is the mass of the $W$-boson, $a$ is a constant, and $E_{sph}$, given by
\begin{equation}
 E_{sph}=(2m_W(T)/\alpha_W)B(m_H/m_W(T))\ ,
\end{equation}
is the energy of the sphaleron, where $m_H$ is the mass of the Higgs boson, $\alpha_w$ is the weak coupling constant, and the function $B$ takes values between $1.56$ and $2.72$. So, one has that $E_{sph}\sim 10$ TeV, and therefore, $\Gamma_{sph}$ is usually small at finite $T$ just below the EW phase transition at $T\sim 100$ GeV. Above the EWPT, the Higgs boson is still in its symmetric vacuum, no symmetry breaking has occurred yet and the mass of the $W$-boson is zero. From the dimensional analysis, it can be argued that $\Gamma_{sph}/V\sim \kappa \alpha_W^4 T^4$ and lattice calculations show that it is $\Gamma_{sph}=(25.4\pm 2.0)\alpha_W^5T^4$. Taking the thermal volume ($V=T^{-3}$), it can be found that $\Gamma_{s}\simeq 10^{-6}$ T. Thus, a comparison of this rate with the expansion rate of the Universe leads to $T_{\ast}\simeq 10^6g_{\ast}^{-1/2}(T)m_{Pl}\simeq 10^{12}$ GeV, and so the sphaleron processes enter in equilibrium when the temperature drops below this value. Any net baryon number and lepton number are washed out by sphaleron processes if no net $B-L$ exists when the sphaleron processes finally enter in equilibrium.

\subsection{The EWB Mechanism}

For EWB to happen, the initial conditions considered are a hot, radiation-dominated early Universe containing zero net baryon charge in which the full $SU(2)_L \times U(1)_Y$ electroweak symmetry is manifest~\cite{Kirzhnits:1972iw, Kirzhnits:1972iw2, L. Dolan, Morrissey:2012db, S. Weinberg}. When the temperature in the Universe drops below the electroweak scale, the electroweak symmetry spontaneously breaks due to the Higgs field settling into a vacuum state. It is during this phase transition that the baryogenesis process in this theory occurs.
Theoretically, this transition proceeds when bubbles of the broken phase nucleate within the surrounding plasma in the symmetric phase, and these bubbles expand, collide and coalesce until only the broken phase remains~\cite{Morrissey:2012db,Cohen:1993nk}. 
The ways in which baryons may be produced as a bubble wall that sweeps through space is separated into two categories~\cite{Riotto:1999yt}: \emph{(i) Nonlocal baryogenesis}---Particles engage with the boundary of the bubble in a manner inconsistent with CP symmetry, introducing an asymmetry in a quantum number distinct from the baryon number in regions of the unaltered phase that are distant from the boundary. Processes that violate baryon number transform the pre-existing asymmetry into a baryon number asymmetry, resulting in the generation of baryons (see~\cite{Joyce:1994zn,Joyce:1994zt} for more details). \emph{(ii) Local baryogenesis}---Baryons are produced when the baryon number and CP violating processes occur together in proximity of the walls.

In general, both of them will occur, and so the baryon asymmetry will be the sum of that generated by the two processes. However, there are conditions that may establish which one dominates. In fact, if the speed of the wall is greater than the sound speed in the plasma, then local baryogenesis dominates. Otherwise, nonlocal baryogenesis is usually more efficient~\cite{Riotto:1999yt}. 
So, summarizing, there are two different limiting regimes for the nonlocal baryogenesis depending on which particular fermionic species are under consideration: 
\begin{enumerate}
\item \emph{The thin wall regime or nonadiabatic regime}: ${L_{\omega}}/ {l}\leq 1$;
\item \emph{The thick wall regime or adiabatic regime}: ${L_{\omega}}/ {l}\geq 1$;
\end{enumerate} 
where $l$ is the mean free path of the fermions being considered, and $L_{\omega}$ is the thickness of the wall. Both of them make use of a master equation of the type 
\begin{equation}
\frac{dN_B}{dt}=-n_f\frac{\Gamma_{\textrm{sph}}(T)}{T}\Delta F \ , \label{masterS}
\end{equation}
where $\Delta F$ is the free energy difference between two neighboring minima and is usually expressed in terms of chemical potentials.
However, to be more precise, the thick wall regime $\Delta F$ is not related to a true chemical potential because the plasma within the bubble wall is not in equilibrium, rather it can be considered as being in quasistatic thermal equilibrium with a classical time-dependent field. Nevertheless, the deviation from the chemical equilibrium may be treated by introducing chemical potentials to the slowly \mbox{varying quantities.}

\subsection{Necessary Conditions for a Successful EWB: The Sphaleron Bound}

There are two requirements that any model of EWB proceeding by the bubble wall scenario must satisfy in order to be successful. On the one hand, if sphaleron processes need to be effective at the symmetric phase so that sphaleron transitions may be in equilibrium and baryon number violation may be completely efficient outside the bubble, on the other hand, it is necessary to ensure that after the time at which the phase transition is complete, the sphaleron processes are indeed ineffective so that the previously generated asymmetry does not become diluted.

The satisfaction of the first condition can be inspected through comparison of the rate for baryon non-conserving processes at the symmetric phase with the expansion rate. Using the thermal rate per unit volume of sphaleron events given by~\cite{Riotto:1999yt}
\begin{equation}\label{rateunbroken}
\Gamma_{\textrm{sph}}^{\textrm{sym}}(T)=\kappa (\alpha_W T)^4\ ,
\end{equation}
the former can be written as
\begin{equation}
V_B^{\textrm{sym}}=\frac{\mathcal{M}}{T^3}\Gamma_{\textrm{sph}}^{\textrm{sym}}=\mathcal{M}\kappa\alpha_W^4T\,, 
\end{equation}
where $\mathcal{M}=13/2n_f$

The prefactor $\kappa$ was calculated in~\cite{1Beds}, where it was shown that it has an extra dependence on $\alpha_W$ so that  $\kappa\approx 25\alpha_W$. Using this result, it is possible to find that the equilibrium condition for sphaleron transitions at the symmetric phase  ($V_{B}^{\textrm{sym}}(T)>H(T)$) is \mbox{satisfied whenever}
\begin{equation}
T< 7.53\mathcal{M}\alpha_W^5\frac{m_{Pl}}{\sqrt{g_{\ast}}} \ ,
\end{equation}
and it is therefore useful to define
\begin{equation}\label{equilibrium sphaleron GRB}
T_{\textrm{sph}}\equiv 7.53\mathcal{M}\alpha_W^5\frac{m_{Pl}}{\sqrt{g_{\ast}}}\ ,
\end{equation}
where $T_{\textrm{sph}}$ is the temperature marking the threshold point at which equilibrium for sphaleron transitions is established. For the standard model, where $g_{\ast}=g_{SM}=106.75$ and $g_{SM}$ are the standard model degrees of freedom, we have $T_{GR}^{\textrm{sph}}\simeq 7.5\times 10^{12}$ GeV. Thus, it is possible to infer that the temperature at which sphaleron processes reach equilibrium is very high and does not represent a source of concern. Relative to the second requirement, things are somewhat different, and it can be regarded as a further criterion that must be satisfied for the bubble wall scenario to be successful. Thus, to establish the bound at which this criterion is satisfied, it is useful to begin by writing the master equation, given by Equation \eqref{masterS}, as
\begin{equation}
\frac{dN_{B}}{dt}=-n_f^2\frac{\Gamma_{\textrm{sph}}}{T}\frac{\partial F}{\partial (N_B+N_L)}=-n_{f}^2\frac{\Gamma_{\textrm{sph}}}{T}(\mu_B+\mu_L)\label{masterS2} \ ,
\end{equation}
where $\mu_B$ and $\mu_L$ are the chemical potentials for the baryon number $B$ and lepton number $L$, respectively. Assuming now that $B-L=0$, $\mu_B+\mu_L$ can be easily computed (see~\cite{Arnold:1987mh} for details), and is given by
\begin{equation}
\mu_B+\mu_L=\frac{13}{2}\frac{N_B}{n_f}T^{-2} \ ,
\end{equation}
and therefore, after substitution of this result into the master equation given above, i.e.,\linebreak \mbox{Equation \eqref{masterS2}}, one can formally integrate it to obtain
\begin{equation}\label{intS}
\frac{N_{B_f}}{N_{B_C}}=\exp\left[-\frac{13}{2}n_f\int_{t_{Cr}}^{t_f}\frac{\Gamma_{\textrm{sph}}(T(t'))}{T(t')^3}dt'\right] \ ,
\end{equation}
where $t_{Cr}$ should be in this case interpreted as the time at which the bubble nucleation begins, and $t_f$ as the time at which the processes stop being effective so that $N_{B_f}/N_{B_C}$ can be understood as an expression for the dilution of the baryon asymmetry, and thus, $B_f$ represents the final asymmetry that should be preserved until today. On the other hand, the thermal rate per unit time and unit volume for fluctuations between neighboring minima given by~\cite{Riotto:1999yt}
\begin{equation}\label{brokenrate}
\Gamma_{\textrm{sph}}^{\textrm{bro}}(T)= \mu \left(\frac{M_W}{\alpha_W T}\right)^{3}M_{W}^4\exp\left(-\frac{E_{\textrm{sph}}(T)}{T}\right) \ ,
\end{equation}
can be written in terms of the sphaleron energy as 
\begin{align}
\Gamma_{\textrm{sph}}&=2\pi^4\mu T^4\left(\frac{\alpha_W}{4\pi}\right)^4\left(\frac{E_{\textrm{sph}}(T)}{BT}\right)^{7}\exp\left[-\frac{E_{\textrm{sph}}(T)}{T}\right]
    \nonumber\\
&\simeq 2.8\times10^5\kappa T^4\left(\frac{\alpha_W}{4\pi}\right)^4\left(\frac{E_{\textrm{sph}}(T)}{TB}\right)^{7}\exp\left[-\frac{E_{\textrm{sph}}(T)}{T}\right] \ ,
\end{align}  
where the approximation taken in the last step can be found in reference~\cite{3Beds}. The prefactor was calculated in~\cite{4Beds} and found to be within the range $10^{-4}\lesssim\kappa\lesssim0.1$. For simplicity, it is useful to write the following results after defining the rate for baryon non-conserving processes as
\begin{equation}
V_B(T)\equiv \mathcal{M}\frac{\Gamma_{\textrm{sph}}}{T^3} \ .
\end{equation}

After the substitution of this definition in the integral Equation \eqref{intS} and taking some assumptions, one obtains a simple expression. More specifically, as the dominant contribution for the integral Equation \eqref{intS} comes from temperatures very close to $T_{Cr}$, for lower temperatures the transitions are highly suppressed by the exponential factor, and one can approximate the integral to the value of the integrand at $T=T_{Cr}$: 
\begin{equation}\label{int2S}
\ln\left(\frac{N_{B_f}}{N_{B_C}}\right)=-V_B(T_C)(t_f-t_C) \ .
\end{equation}

Defining now the dilution as $\mathcal{S}\equiv N_{B_f}/N_{B_C}=(n_{B_f}/s_0)/(n_{B_C}/s_C)$, where $s_0$ is the entropy at the present day, $s_C\equiv s(t_C)$ is the entropy at the time of the phase transition and $n_{B_f}=n_{B}(t_f)$ is the baryon-to-entropy density ratio observed at the present epoch, one has that the maximum dilution that can occur and yet preserve the observed baryon density is given by $\mathcal{S}_{max}\sim 7\times10^{-11} s_C /n_{B_C}$. In other words, the value of the baryon asymmetry, which can be produced at the electroweak phase transition $n_B/s_c$, is still an open question. However, it is generically difficult to produce a large asymmetry and, independently of the model, it is very unlikely to be greater than $10^6$. That way, one can reasonably write the bound $N_{B_C} \lesssim 10^6$. This means that to avoid the washout of the baryon asymmetry, one must fulfill the requirement  $S \gtrsim 10^{-5}$. 

In standard cosmology, the time--temperature relation for a radiation Universe is given by $t=0.301g_{\ast}^{-1} m_{Pl} T^{-2}$; thus, 
\begin{eqnarray}
t_{f}-t_{Cr} & = & 0.301m_{Pl}\left[g_{\ast} (T_{Cr})^{-1} T_{Cr}^{-2}-g_{\ast}(T_F)^{-1}T_f^{-2}\right] 
    \nonumber \\
&\approx & 0.301 m_{Pl} T_{Cr}^{-2} / g_{\ast}(T_{Cr}) \ ,
\end{eqnarray}
where $T_f\approx 0$ can be understood as taking the integration until $t_f=\infty$. So, using the relation between $\phi(T_{Cr})$ and $E_{\textrm{sph}}(T_{Cr})$~\cite{3Beds}
\begin{equation}
\frac{\phi(T_{Cr})}{T_{Cr}}=\frac{g}{4\pi B}\frac{E_{\textrm{sph}}(T_{Cr})}{T_{Cr}} \ ,
\end{equation}
assuming $t_{Cr}-t_f=0.301m_{Pl}T_{Cr}^{-2}/g_{\ast Cr}$, where $g_{\ast Cr}\equiv g_{\ast}(T_{Cr})$, and imposing the dilution condition $\mathcal{S}\gtrsim 10^{-5}$, it is possible to write the sphaleron bound as 
\begin{equation} \label{boundS}
\sqrt{\frac{4\pi}{\alpha_W}}\frac{\phi_{Cr}}{T_{Cr}}B-7\ln\left(\sqrt{\frac{4\pi}{\alpha_W}}\frac{\phi_{Cr}}{T_{Cr}}B\right)
\gtrsim
\ln\left[\frac{2.8\times10^5\kappa\mathcal{M}}{5\ln10\times B^7}\frac{}{}\left(\frac{\alpha_W}{4\pi}\right)^4\right]+\ln\left(\frac{T_{Cr}}{2H_{Cr}}\right) \ ,
\end{equation}
where $\phi_{Cr}\equiv\phi(T_{Cr})$, $H_{Cr}\equiv H(T_{Cr})$ is the standard cosmology expansion rate for a radiation Universe at the time of the phase transition, and $B$ is a function of the Higgs mass $m_H$. For the standard model, $B$ is approximately given by~\cite{3Beds}
\begin{equation}
B\left(\frac{m_H}{m_W}\right)\simeq 1.58+0.32\left(\frac{m_H}{m_W}\right)-0.05\left(\frac{m_H}{m_W}\right)^2\label{BS} \ ,
\end{equation}
an approximation valid for values of the Higgs mass in the range $25~\textrm{GeV}\leq m_H\leq 250~\textrm{GeV}$. This parameter ranges exactly from $B(0)=1.52$ to $B(\infty)=2.70$~\cite{Manton,Klinkhamer,Brihaye,Kunz}. If one uses now $T_{Cr}=10^2$ GeV, $n_{f}=3$, $\alpha_W=0.0336$, $\kappa=0.1$ and naively $B=1.9$, the sphaleron-bound Equation \eqref{boundS} can be solved numerically
\begin{equation}
\frac{\phi_{Cr}}{T_{Cr}}\gtrsim 1.22 \,.
\end{equation}

\subsection{Departure from Equilibrium}

At electroweak temperatures, the equilibrium description of the phenomena is very accurate since the expansion rate of the Universe is small compared to the rate of baryon number-violating processes. Thus, for baryogenesis to happen, it needs the aid of phase transitions. Thus, the central point is to understand if the electroweak phase transition (EWPT) is a first- or second-order phase transition. The solution to this problem is given by the relation of the phase transitions and the expansion rate of the Universe. 
As the Universe cools, at the electroweak scale, the EWPT dictates the possible effects of interest to the mechanisms for baryogenesis. If the EWPT is of the first order, then the nucleation of bubbles characterized by a non-zero Higgs field vacuum expectation value (vev) occurs within the symmetric vacuum. During their expansion, particles within the plasma interact with the dynamic phase interface, called the bubble wall, in a way that violates CP symmetry. This violation leads to the generation of a CP-asymmetric particle flux into the symmetric phase. The resulting CP asymmetry is then transmuted into a baryon asymmetry through sphaleron processes that violate the baryon number, taking place close to the \mbox{advancing wall. }

As the phase interface advances, baryons transition into the broken phase, where the inefficiency of baryon violation leads to the freezing out of the baryon asymmetry (see~\cite{Servant} for more details). As exposed in~\cite{Prokopec}, this mechanism can be seen as ``broken'' and ``unbroken'' phases separated by a potential barrier, which decreases as the Universe cools. When the energy barrier is low enough, the transition advances through the formation and expansion of true vacuum bubbles. This shift from equilibrium happens rapidly compared to the overall expansion timescales. Many proposed ways of creating baryon asymmetry at the electroweak scale rely on this sudden departure from  equilibrium, using the interaction between the plasma and advancing walls to generate the baryon asymmetry. In second-order transitions, this effect is absent. The evolution proceeds smoothly, and the departure from equilibrium is directly influenced by the expansion rate. Thus, this leads to the conclusion that anything other than a first-order phase transition is not conducive to baryogenesis at the electroweak scale. This conclusion reinforces how the expansion rate of the Universe has a fundamental role in the construction of feasible theories of baryogenesis due to the connection with the departure from equilibrium.

\section{GUT Baryogenesis}\label{sec:GUTb}

The goal of Grand Unification Theories (GUTs) is to unify the strong, weak and electromagnetic interactions and quarks and leptons within the framework of a gauge field theory based upon a non-Abelian symmetry group~\cite{Georgi, Mohapatra,Fritzsch,Croon:2019kpe}. A general feature of GUTs is that leptons and baryons are placed in the same multiplets and thus mixed under gauge transformations so that baryon and lepton numbers are not individually conserved. This allows baryon non-conserving interactions, such as $e^{-}+d\leftrightarrow \bar{u}+\bar{u}$, or decays of superheavy gauge (or Higgs) bosons, such as $X\rightarrow e^{-}+d +\bar{u}+\bar{mu}$. To illustrate more carefully how baryon violation proceeds, one can consider that $X$, $\bar{X}$, $q$, $\bar{q}$, $l$, $\bar{l}$ denote respectively an arbitrary superheavy gauge, an arbitrary superheavy gauge antiboson, an arbitrary quark, an arbitrary antiquark, an arbitrary lepton and an abitrary antilepton. As will be seen, since the superheavy bosons decays are the most important processes for baryogenesis, the focus of the next discussion will be on this topic. 

The $X$ bosons have decay modes of the type $X\rightarrow ql$ and  $X\rightarrow\bar{q}\bar{q}$, and the antibosons $\bar{X}$ have decay modes of type $\bar{X}\rightarrow\bar{q}\bar{l}$ and $\bar{X}\rightarrow qq$ (for instance, see~\cite{3Bgi}). One verifies that the decay of a superheavy gauge boson clearly violates baryon number, and thus satisfying the Sakharov's first criterion.
More precisely, if one denotes the branching ratios of each mode of the $X$ decay described above as $r$, $1-r$, $\bar{r}$ and $1-\bar{r}$ respectively, then the mean net baryon number produced by the decay of a boson $X$ is $B_{X}=r/3-2(1-r)/3$, and the mean net baryon number produced by the decay of an antiboson $\bar{X}$ is $B_{\bar{X}}=-\bar{r}/3+2(1-\bar{r})/3$. Therefore, the mean net baryon number produced by the decay of an $X$, $\bar{X}$ pair is just $\varepsilon=B_{x}+B_{\bar{x}}=r-\bar{r}$. However, if one considers, for instance, the channels of decay $X\rightarrow \bar{q}\bar{l}(1)$ and denotes the parity (P) of the state  $(1)$  by ($\uparrow$) or ($\downarrow$), one can write the following transformation properties: 
\begin{eqnarray}
\begin{array}{ll}
\textrm{Under CPT}:&\Gamma(X\rightarrow 1 \uparrow )=\Gamma(\bar{1} \downarrow\rightarrow \bar{X}),\\
\textrm{Under CP}:&\Gamma(X\rightarrow 1 \uparrow )=\Gamma(\bar{X}\rightarrow\bar{1} \downarrow),\\
\textrm{Under C}:&\Gamma(X\rightarrow 1 \uparrow )=\Gamma(\bar{X}\rightarrow\bar{1} \uparrow),
\end{array}
\end{eqnarray}
where $\Gamma$ is to be understood as the $X$ decay width to a state $ql$. Therefore, one concludes that: 
\begin{equation}
\varepsilon=r-\bar{r}=\Gamma(X\rightarrow 1\uparrow)+\Gamma(X\rightarrow 1\downarrow)-\Gamma(\bar{X}\rightarrow \bar{1}\uparrow)-\Gamma(\bar{1}\downarrow)\label{eps1}\,.
\end{equation}

This result holds independently of the choice for state $(1)$. On the other hand, it can be seen that if $C$ or $CP$ are good symmetries, then $\epsilon=0$. So, as expected, two more requisites must be fulfilled: $C$ and $CP$ violation, and, of course, the departure from equilibrium. These theories can involve a substantial discussion of CP violation, which lies outside the scope of this review paper. The focus will be directed toward the departure from equilibrium and the so-called Non-Inflationary GUT-Baryogenesis Mechanism.

\subsection{Departure from Equilibrium}

The third Sakharov criterion is satisfied if the expansion of the Universe is faster than the particle interaction rates. In other words, the crucial issue is how this expansion rate compares with the rate of the relevant microphysical processes. One considers that at some early times when the bosons that mediate baryon non-conserving processes are still relativistic, they are present in their equilibrium numbers, that is $n_{X}=n_{\bar{X}}=(g_{x}/g_{\gamma})n_{\gamma}=(\zeta(3)/\pi^2)g_X T^3$, where the first equality accounts for symmetric initial conditions. In other words, it stands for an equal number of bosons $X$ and bosons $\bar{X}$.  However, as the temperature drops, the numbers of bosons $X$, $\bar{X}$ will only remain within their equilibrium values if the interactions that create and annihilate the bosons (decay, annihilation and their inverse processes) occur rapidly on the expansion times scale $\Gamma\gtrsim H$, or reversely. If these processes occur slowly relative to the expansion time scale ($\Gamma\lesssim H$), they decouple from the surrounding medium and so their number freezes and it is always comparable to the number of photons. 

If the bosons are indeed massive enough, the reactions that decrease the number of $X$ and $\bar{X}$, that is, annihilation and decay processes, are ineffective even for $T < m_X$ so that their actual number will strongly depart from their equilibrium values. The aforementioned processes are then unable to reduce the bosons $X$ and $\bar{X}$ in order for them to follow their exponential depletion of the equilibrium values, and they will become overabundant, that is, $n_X\simeq n_{\bar{X}}\simeq n_{\gamma}\gg n_X^{\textrm{eq}}=n_{\bar{X}}^{\textrm{eq}}=(m_XT)^{3/2}\exp(-m/T)$. This overabundance corresponds to the departure from thermal equilibrium needed to generate a nonvanishing baryon asymmetry when the heavy states $X$ undergo $B$- and $CP$-violating decays. Considering that the heavy particles decay through renormalizable operators, the out-of-equilibrium condition requires very heavy states: $m_X\gtrsim 10^{15}-10^{16}$ GeV and $m_X\gtrsim 10^{10}- 10^{16}$ GeV for gauge and scalar bosons, respectively~\cite{Riotto:1999yt,Mark Trodden,Kolb:1990vq}. The annihilation process is self quenching since $\Gamma_{\textrm{ann}}\propto n_X$ and the decay process is important for maintaining the equilibrium numbers of $X$, $\bar{X}$ bosons. For simplicity then, we will ignore the annihilation process.

\subsection{Non-Inflationary GUT-Baryogenesis Mechanism}

Whenever inflation is not incorporated, it is generally assumed that the Universe is filled with a hot soup containing all fundamental particles in thermal equilibrium, including the supermassive bosons that mediate baryon-nonconserving processes. This is a sensitive point of the model since as shown in Section \ref{Equilibrium in the Expanding Universe}, at temperatures slightly higher than the GUT scale ($T\sim 10^{16}$ GeV) around which the baryogenesis should occur, none of the known interactions, nor those arising for the GUTs, are capable of establishing or maintaining thermal equilibrium. In general, there are several processes involving the superheavy bosons that may influence the final value of the baryon asymmetry. Up to the need to consider the second order on $\alpha_S$, these comprise decays and inverse decays of superheavy bosons (of order $\alpha_S$), baryon-nonconserving fermion collisions (BNC-scattering processes) and Compton-type and annihilation-like reactions (both of order $\alpha_S^2$). Most of the time, second-order processes are not important and, so, it is expected that higher-order processes are negligible~\cite{2Bgi}. In any case, within the processes of order $\alpha_S^2$, the BNC-scattering processes are the only ones which, in most situations, may have some relevance. Processes other than the decay will only have the effect of damping any generation of asymmetry that might be produced by the decays themselves. 

The success of these two damping processes in opposing the decays depends on their relative rates to the expansion rate of the Universe. Thus, if only the decay, inverse decay and BNC-scattering processes are taken into account, then the departure from equilibrium can be quantified in terms of the rates for each of these processes. Formal calculations of those rates can be found in reference~\cite{2Bgi}. They are given by
\begin{eqnarray}
\Gamma_D\simeq\alpha_X \frac{m_X^2}{\sqrt{T^{2}+m_X^2}}\simeq \Bigl\{\begin{array}{ll}\alpha_Xm_X^2/T&\textrm{if}~T\gtrsim m_X\\
\alpha_X m_X&\textrm{if}~T\lesssim m_X\end{array} \ , \\
\Gamma_{ID}\simeq\Gamma_{D} X^{eq}\simeq2\Gamma_D \Bigl\{\begin{array}{ll} 1&\textrm{if}~ T\gtrsim m_X \\
\sqrt{\pi/8}(m_X/T)^{3/2}\exp(-m_X/T)&\textrm{if}~  T\lesssim m_X \end{array} \ ,\\
\Gamma_S\simeq A\alpha_X^2~\frac{T^5}{(T^2+m_X^2)^2}\simeq A\alpha_X^2 \Bigl\{\begin{array}{ll}m_X T&\textrm{if}~T\gtrsim m_X\\ T^{5}/m_S^4&\textrm{if}~T\lesssim m_X \end{array} \ ,
\end{eqnarray}
where $X^{\textrm{eq}}$ is the equilibrium number of bosons $X$, $\gamma_{ID}$ is assigned to the inverse decay processes rate, $\gamma_S$ is assigned to the BNC-scattering processes rate, and $A$ is a numerical factor of order few $\times 10^3$, which accounts for the number of scattering channels, etc. For purposes of baryogenesis, the most important rate is the decay rate since the decays are the mechanism that regulates the number of $X$, $\bar{X}$ bosons (notice that indeed the decays are the dominant means of reducing the number of superheavy bosons since the annihilation is of order $\alpha^2$) since these are indeed the key particles that mediate baryon nonconservation. It is then useful to define a quantity  
\begin{equation}   
K\equiv \left(\frac{\Gamma_D}{2 H}\right)_{T=m_X}=3.68 \times10^{18}\frac{\alpha_X}{\sqrt{g_{\ast S}}}\frac{\textrm{GeV}}{m_X}\label{K}\,, 
\end{equation}
which is proportional to the effectiveness of the decays and inverse decays ($\Gamma_D/H,~\Gamma_{ID}$, respectively) and to $\alpha^{-1}$ times the effectiveness of the BNC-scattering processes ($\Gamma_{S}/H\alpha$) at the crucial time at which the bosons $X$ become nonrelativistic ($m_X=T$) and, therefore, are forced to be rapidly reduced in number if they are to stay in equilibrium.

\section{Scalar--Tensor Theories}\label{sec:STT}

\subsection{Action and Field Equations}
When extending Einstein's GR, albeit preserving its metric framework~\cite{Will:2018bme}, the
Scalar--Tensor Theories (STT) 
occupy a central place~\cite{Avelino:2016lpj}. They are characterized by the inclusion of a 
scalar field mediating the gravitational interaction, in addition to the metric tensor field. 
The archetypal STT theory is Brans--Dicke theory 
put forward in 1961~\cite{Brans}, and soon after generalized to theories with arbitrary coupling parameters by Bergmann~\cite{Bergmann}, Wagoner~\cite{Wagoner} and Nordtvedt~\cite{Nordtvedt}. The fundamental action for these theories can be cast in the \mbox{following form:}
\begin{equation}\label{actionSTT}
S=\frac{1}{16\pi}\int\sqrt{-g}\,d^4x\Big[\phi [R-2\lambda(\phi)]-\frac{\omega(\phi)}{\phi}g^{\mu \nu}\partial_\mu\phi\partial_\nu \phi\Big]+S_M \ ,
\end{equation}
where $\omega(\phi)$ is a dimensionless function of $\phi$ which calibrates the coupling between the scalar field and gravity (in future references, $\omega(\phi)$ will be termed as the \emph{coupling parameter}) and $\lambda(\phi)$ is another function of the scalar field which can be interpreted both as a potential for $\phi$, and as a cosmological parameter.
Therefore, the scalar field $\phi$ plays the role that is associated with
the gravitational constant $G_N$ in Einstein's GR. 
And, due to the presence of a non-vanishing kinetic term ($g^{\mu\nu}\phi_{,\mu}\phi_{,\nu}$) in the action, it is 
no longer a constant but instead a dynamical variable with dimensions of squared mass.  The STT may be perceived as arising from the low-energy limit of superstrings theory since the Brans--Dicke field may be related to the dilaton~\cite{Brandenberger:2023ver,Green:2012oqa}.

Thus, unlike what happens in GR, the scalar--tensor action describes a gravitational theory in which two fields mediate the gravitational interaction, the metric tensor $g_{\mu\nu}$ and a field $\phi$. Furthermore, by applying the variational principle to this action, one obtains two equations. One modified field equation from the variation concerning the metric $\delta g_{\mu\nu}$, and an equation of motion for the scalar field, from the variation concerning the scalar $\delta \phi$. 

The equations mentioned are given by
\begin{eqnarray}
    R_{\mu\nu}-\frac{1}{2}g_{\mu\nu}R=\lambda(\phi)g_{\mu\nu}+\frac{\omega(\phi)}{\phi^2}\Big(\partial_\mu \phi \partial_\nu \phi -\frac{1}{2}g_{\mu\nu}\partial^\alpha \phi \partial_\alpha \phi \Big)
        \nonumber \\
    +\frac{1}{\phi}(\nabla_\mu \partial_\nu\phi-g_{\mu\nu}\Box\phi)+8\pi G_N\frac{T_{\mu\nu}}{\phi} \ ,
    \label{field equations1 STT}
\end{eqnarray}
\begin{equation}\label{KG-STT}
    (2\omega(\phi)+3)\Box \phi=8\pi G_N T-\omega'(\phi) \partial_\alpha\phi \partial^\alpha \phi-2\phi^2\lambda'(\phi)+4\phi\lambda(\phi) \ ,
\end{equation}
respectively, where Equation \eqref{KG-STT} can be seen as a modified classical Klein--Gordon equation, being obtained from the substitution of the trace of Equation (\ref{field equations1 STT}), which yields $R=-\phi^{-1}[8\pi G_N T-\phi^{-1}\omega(\phi)\partial_\alpha \phi\partial^\alpha \phi-3\Box\phi]-4\lambda(\phi)$.

\subsection{Scalar--Tensor Cosmology: 3-Epoch Model}

In order to obtain the cosmological solutions for STT~\cite{Barrow:1994nx}, we use the field\linebreak Equations \eqref{field equations1 STT} and \eqref{KG-STT}, assuming as before a Universe described by the FLRW metric and assuming a perfect fluid matter content described by the barotropic equation of state \eqref{baritropic PFluid}. Thus, disregarding the term $\lambda(\phi)$, one obtains 
\begin{align}
\left(\frac{\dot{a}}{a}\right)^2+\frac{\dot{a}}{a}\frac{\dot{\phi}}{\phi}+\frac{k}{a^2}-\frac{\omega(\phi)}{6}\frac{\dot{\phi}^2}{\phi^2}=&\frac{8\pi}{3}\frac{\rho}{\phi},\label{STTFRW1}\\
\frac{\ddot{a}}{a}-\frac{\dot{a}}{a}\frac{\dot{\phi}}{\phi}-\frac{1}{2}\frac{\dot{\omega}(\phi)}{2\omega(\phi)+3}\frac{\dot{\phi}}{\phi}=&-\frac{8\pi}{3}\frac{\rho}{\phi}\frac{(3\gamma-2)\,\omega(\phi)+3}{2\omega(\phi)+3}-\frac{\omega(\phi)}{3}\frac{\dot{\phi}}{\phi},\label{STTFRW2}\\
\ddot{\phi}+\left[3\frac{\dot{a}}{a}+\frac{\dot{\omega}(\phi)}{2\omega(\phi)+3}\right]\dot{\phi}=&\frac{8\pi\rho}{2\omega(\phi)+3}(4-3\gamma)\label{STTFRW3}\;.
\end{align}

We consider a 3-epoch model explored in~\cite{Mimoso:1992vr,Mimoso:1994}, and we refer the reader to latter references for further 
details of the model. 
As a first step, the model is built on the assumptions that the Universe is well described by an FLRW metric and by an energy--momentum of the form of Equation \eqref{Perfect_fluid}, that is, it assumes a perfect fluid as the matter content of the Universe; the barotropic equation of state is given 
by Equation \eqref{baritropic PFluid}, the curvature can be disregarded, that is, $k=0$ is assumed, and $\lambda(\phi)$ is vanishing. The model has three distinct epochs characterized by the domination of different types of fluid and different values of the coupling parameter $\omega$, which are incorporated as the main stages of the evolutionary Universe. 

In order to obtain the results according to which the effects at a given instant of a general scalar--tensor theory can be reduced to those of the Jordan--Brans--Dicke (JBD) theory, it is useful to develop the STT field Equations \eqref{STTFRW1}--\eqref{STTFRW2}, that is, the scalar--tensor equations where the coupling parameter $\omega$ is allowed to depend on the scalar field $\phi$ but where the potential $\lambda$ is vanishing. To this effect, it is useful to introduce the conformal time $\eta$ given by the differential relation
\begin{equation}
dt=ad \tau \,,
\end{equation}
the variables~\cite{Mimoso:1994wn}
\begin{align}
X&\equiv \Phi a^2,\\
Y&\equiv \int\,\sqrt{\frac{2\omega+3}{3}}\,\frac{d\Phi}{\Phi},\label{redefined_omega}
\end{align}
and the relation between $\rho(t)$ and $a(t)$
\begin{equation}
\rho(t)=\rho_0\left(\frac{a}{a_0}\right)^{-3\gamma}\,,
\end{equation}
which comes directly from the integration of the equation for the conservation of energy Equation \eqref{Conservação} using the barotropic equation of state, given by Equation \eqref{baritropic PFluid}. The field Equations \eqref{STTFRW1}--\eqref{STTFRW3} can be rewritten as
\begin{align}
&(X')^2+4k\,X^2-(Y'X)^2=4M\,X\,\left(\frac{X}{\Phi}\right)^{\frac{4-3\gamma}{2}},\\
&X''+4k\,X=3M(2-\gamma)\,\left(\frac{X}{\Phi}\right)^{\frac{4-3\gamma}{2}},\\
&(Y'X)'=M(4-3\gamma)\,\sqrt{\frac{3}{2\omega+3}}\,\left(\frac{X}{\Phi}\right)^{\frac{4-3\gamma}{2}}\ ,
\end{align}
where the prime denotes now differentiation with respect to $\eta$, and $M\equiv 8\pi\rho_0/3$ with $\rho_0$ as an arbitrary initial condition for the energy density.

There are two  fundamental features of the STT cosmological solutions that are relevant to the analysis of  the interplay between quantum interactions and the expansion rate of the Universe during the phases that are relevant for baryogenesis. On the one hand, the latter processes under consideration take place during the radiation epoch, whose early stages of expansion are dominated by the scalar field energy density~\cite{Barrow:1994nx,Mimoso:1994wn}. On the other hand, Equation~(\ref{redefined_omega}), which is valid during the radiation epoch, enables one to define an average value of the coupling $<\omega>$ during any time partition of the radiation epoch, even though $\omega$ may be strictly increasing. Indeed, between any time interval $[t_i,t_j]$, we define
\begin{equation}
\sqrt{\frac{2<\omega>_{ij}+3}{3}} = \frac{\int_{t_i}^{t_j}\,\sqrt{\frac{2\omega+3}{3}}\,\frac{d\Phi}{\Phi}}{\int_{t_i}^{t_j}\,\frac{d\Phi}{\Phi}}\; ,
\end{equation}
and hence we can treat a general STT as a Brans--Dicke theory with  a constant coupling $<\omega>_{ij}$, averaged in the time interval.  

We choose to divide the radiation epoch in two intervals corresponding to a first stage of JBD scalar field domination, followed by a second stage during which radiation and ultra-relativistic matter dominate over the scalar field. Finally, the radiation epoch is followed by a matter-dominated phase $p\sim 0$. This is encapsulated in the following $3$-epoch solution of the general JBD solution for a flat FLRW Universe:
\begin{align}
0\leq t\leq t_c\,:~~a(t)=&A_1 t^q , \\
t_c\leq t\leq t_{eq}\,:~~a(t)=&A_2(t-t_{20})^{1/2} , \\
t_{eq}\leq t_0\leq t_c\,:~~a(t)=&A_3(t-t_{30})^{\frac{2\omega+2}{3\omega+4}}\,,
\end{align}
where $q$ is given by~\cite{OHanlon:1972ysn}
\begin{equation}
q\equiv\frac{\omega}{3\Big(\omega+1\mp\sqrt{\frac{2\omega+3}{3}}\Big)} \ ,\label{qformula}
\end{equation}
and $A_1$, $A_2$, $A_3$, $t_{20}$ and $t_{30}$ are constants which have to satisfy junction conditions guaranteeing the smoothness of the solution. Imposing the scale factor and its first derivative to be continuous at $t=t_c$ and $t=t_{eq}$ as smoothness conditions leads to 
\begin{align}
A_2=&\sqrt{2q}\,A_1\,t_c^{q-1/2},\\
t_{20}=&\Big(1-\frac{1}{2q}\Big)\,t_c \ ,
\end{align}
and
\begin{align}
A_3=&\sqrt{2q}\,A_1\,t_c^{q-1/2}\,\Big(\frac{4\omega+4}{3\omega+4}\Big)^{-\frac{2\omega+2}{3\omega+4}}\,(t_{eq}-t_{20})^{\frac{1}{2}-\frac{2\omega+2}{3\omega+4}},\\
t_{30}=&-\frac{\omega}{3\omega+4}\,t_{eq}+t_{20} \,.
\end{align}

Now, in this solution, the BD scalar field evolves as:
\begin{align}
0\leq t\leq t_c\,:~~\phi(t)=&\phi_1\,t^{\frac{1-3q}{3}},\label{phi1}\\
t_c\leq t\leq t_{eq}\,:~~\phi(t)=&\phi_2,\\
t_{eq}\leq t_0\leq t_c\,:~~\phi(t)=&\phi_3\,(t-t_{30})^{\frac{2}{3\omega+4}}\,,\label{phi3}
\end{align}
where $\phi_1$, $\phi_2$ and $\phi_3$ are constants which satisfy continuity the requirements for $\phi$ at $t_c$ and at $t_{eq}$. Note that for the first epoch, that is, for $0\leq t\leq t_c$, $\dot{\phi}/\phi$ is negative for $q>1/3$ and since the choice of the $(-)$ sign (corresponding to a $q_-$ branch) in Equation \eqref{qformula} leads to $1/3<q\leq1$, it is possible to find that the choice of this sign always leads to a decreasing $\phi$, which consequently leads to a $G$ approaching $\infty$ with time. In the case of $q>1$, which corresponds always to the choice of the $(+)$ sign (the $q_+$ branch), this remains true but leads to an accelerated expanding solution associated with the range $-3/2\leq\omega<-4/3$. On the other hand, for $q<1/3$, which corresponds always to the choice of the $(+)$ sign, $\dot{\phi}/\phi$ is positive and therefore $\phi$ increases with time, which means that $G$ approaches zero in the $t\rightarrow\infty$ limit. It is, however, important to mention that for $q<0$, that is, $-4/3<\omega<0$, the scale factor contracts with time.

\section{Scalar--Tensor Baryogenesis}\label{sec:STTbaryogenesis}

In this section, we will briefly explore how a change in the theory of gravity affects the EWB and the GUT baryogenesis. We focus on the main differences that arise in this extended gravity theory with regard to the GR framework.


\subsection{Scalar--Tensor EWB}

A different thermal history of the Universe can have an influence on the EWB. This reasoning is due to the sphalareon bound, which, if not satisfied by a particular model of EWB, excludes it as a viable hypothesis for the explanation of the baryon asymmetry of the Universe (BAU). 

The sphaleron bound is derived assuming a specific expansion rate of the Universe, and it can easily be relaxed if another thermal history of the Universe is assumed. In the STT case, the bound has to be rewritten in terms of the expansion rate for the first or second cosmological epochs.  Nevertheless, the second epoch, due to its similarity to GR, is not expected to have a great influence in relaxing the sphaleron bound. So, because the third epoch happens much later than the time window that is envisaged, we start by writing the STT expansion rate for the first two epochs
in terms of both the temperature and the expansion rate for a radiation Universe in standard cosmology ($H$) 
\begin{eqnarray}
\begin{array}{ll}
\Gamma_{e_1}(T)=\frac{H(T)}{\zeta}\Big(\frac{T_c}{T}\Big)^{\frac{2q-1}{q}}&;~T\geq T_c\\
\Gamma_{e_2}(T)=\frac{H(T)}{\zeta}&;~T_c\geq T\geq T_{eq}
\end{array} \label{Time-T_STT}
\end{eqnarray}
where $H(T)=1.66\sqrt{g_{\ast}}T^{2}/m_{Pl}$ and $\zeta$ is a parameter given by~\cite{Mimoso:1992vr,Mimoso:1994}
\begin{equation}
\zeta=\Big[\Big(\frac{\rho_0^R}{\rho_0^M}\Big)^{\frac{1}{2\omega+2}}\Big] \ ,
\end{equation}
where $\rho_0^R$ and $\rho_0^M$ are the present values of the radiation energy density and matter energy density, respectively.

It becomes clear that the sphaleron bound, given by 
Equation \eqref{boundS}, can be rewritten for the STT as
\begin{eqnarray}\label{boundSTTS}
\sqrt{\frac{4\pi}{\alpha_W}}\frac{\phi_{Cr}}{T_{Cr}}B-7\ln\Big(\sqrt{\frac{4\pi}{\alpha_W}}\frac{\phi_{Cr}}{T_{Cr}}B\Big)\gtrsim \hspace{195pt}\nonumber\\
\Biggl\{\begin{array}{ll}
\ln\Big[\frac{2.8\times10^5\kappa\mathcal{M}}{5\ln10~B^7}\Big(\frac{\alpha_w}{4\pi}\Big)^4\Big]+\ln\Big(\frac{T_{Cr}}{2H_{Cr}}\Big)+\Big(\frac{2q-1}{q}\Big)\ln\Big(\frac{T_{Cr}}{T_c}\Big)+\ln(\zeta) &;~T_{Cr}\geq T_c\label{boundSTTS1}\\
\ln\Big[\frac{2.8\times10^5\kappa\mathcal{M}}{5\ln10~B^7}\Big(\frac{\alpha_w}{4\pi}\Big)^4\Big]+\ln\Big(\frac{T_{Cr}}{2H_{Cr}}\Big)+\ln(\zeta)&;~T_{Cr}\leq T_c 
\end{array} \ .
\end{eqnarray}

Thus, one can see that if one wishes to significantly weaken the sphaleron bound, the phase transition should happen during the first epoch. 

From Equation~(\ref{Time-T_STT}), it is possible to see that $q$ must be smaller $1/2$; otherwise, the expansion rate for the first epoch will have a negative influence over the sphaleron bound, increasing its lower bound rather than decreasing it. The reason behind this is that for $q>1/2$, the expansion rate for the STT first cosmological epoch is lower than the standard GR expansion rate for a radiation-dominated Universe, and so the sphaleron processes at the broken phase will be more effective in damping the asymmetry.

Besides the condition leading to a strong first-order transition, there is another constraint that has to be imposed. To promote an efficient baryon number violation, one has to ensure that sphalerons are at equilibrium in the symmetric phase. This can be achieved by imposing that the rate for baryon non-conserving processes at the broken phase is greater than the expansion rate of the Universe. So, using the thermal rate per unit volume of sphaleron events, Equation \eqref{rateunbroken}, one can write the rate for baryon non-conserving processes at the symmetric phase as
\begin{equation}
V_B^{sym}=\frac{\mathcal{M}}{T^3}\Gamma_{sph}^{sym}=\mathcal{M}\kappa\alpha_W^4T \ .
\end{equation}

The prefactor $\kappa$ was calculated in Ref.~\cite{1Beds}, where it was shown that it has an extra dependence on $\alpha_W$, so that  $\kappa\approx 25\alpha_W$. Using this result, it is possible to explicitly write the rate for baryon non-conserving processes in the broken phase as
\begin{equation}
V_{B}^{sym}=25\mathcal{M}\alpha_W^5T \ .
\end{equation}

Thus, in the standard cosmological scenario, the condition for the sphaleron transitions in equilibrium at the symmetric phase is given by
\begin{equation}
V_{B}^{sym}(T)>H(T)\Leftrightarrow T< 7.53\mathcal{M}\alpha_W^5\frac{m_{Pl}}{\sqrt{g_{\ast}}} \ ,
\end{equation}
and it is therefore useful to define 
\begin{equation}
T_{GR}^{sph}\equiv 7.53\mathcal{M}\alpha_W^5\frac{m_{Pl}}{\sqrt{g_{\ast}}} \ ,
\end{equation}
where $T_{GR}^{sph}$ is the temperature at which the sphaleron transitions start to be in equilibrium. For the standard model, where $g_{\ast}=g_{SM}=106.75$ and $g_{SM}$ are the standard model degrees of freedom, we have $T_{GR}^{sph}\simeq 7.5\times 10^{12}~GeV$.
Similarly, for the first STT cosmological epoch, the condition for the sphaleron transitions in equilibrium at the symmetric phase can easily be written as
\begin{equation}\label{temperatureSTT}
V_{B}^{sym}(T)>\Gamma_{e_1}(T) \qquad \Leftrightarrow \qquad
T<T_c\Big(\frac{T_{GR}^{sph}}{T_c}\zeta\Big)^{\pm\big\vert\frac{q}{1-q}\big\vert} \ ,
\end{equation} 
where the sign $(-)$ stands for $q<0\vee q>1$ and the sign $(+)$ stands for $0<q<1$. Analogously, one can define
\begin{equation}
T_{STT}^{sph}\equiv T_c\Big(\frac{T_{GR}^{sph}}{T_c}\zeta\Big)^{\pm\big\vert\frac{q}{1-q}\big\vert} \ ,
\end{equation}
where $T_{STT}^{sph}$ is the temperature assigned to the crossover between $V_B^{sym}(T)$ and $\Gamma_{e_1}(T)$.

Once this is secured,
it is then important to impose that the phase transition occurs  after the sphalerons processes reach their equilibrium with the symmetric phase, as well as that the nucleosynthesis takes place during the second epoch after the phase transition. 
Under these requirements, one faces the following possibility: the sphaleron processes enter equilibrium in the symmetric phase before the phase transition, and the phase transition occurs before the transition to the second epoch, which occurs before nucleosynthesis. This hypothesis, which may have some relevance for the EWB, can be translated in terms of temperature in the following way:
\begin{eqnarray}
T_{GR}^{sph}>\zeta^{-1}T_{Cr}^{\frac{1-q}{q}}T_c^{\frac{2q-1}{q}}>\zeta^{-1}T_c>\zeta^{-1}T_N^{\frac{1-q}{q}}T_c^{\frac{2q-1}{q}}\,,& \quad \textrm{if}~0<q<1 \ , \label{primeiraqmenorqueEW}\\
T_{GR}^{sph}<\zeta^{-1}T_{Cr}^{\frac{1-q}{q}}T_c^{\frac{2q-1}{q}}<\zeta^{-1}T_c<\zeta^{-1}T_N^{\frac{1-q}{q}}T_c^{\frac{2q-1}{q}}\,,& \quad \textrm{if}~q<0\vee q>1 \ . \label{primeiraqmaiorqueEW}
\end{eqnarray}

To proceed with the analysis, we consider $q < 1/2$ so that the condition \eqref{primeiraqmaiorqueEW} can be recast as
\begin{eqnarray} 
\frac{1}{2}>q>\ln T_c\left[\ln\left(\frac{T_c^2}{\zeta}\frac{T_{Cr}}{T_{GR}^{sph}}\right)\right]^{-1}  \qquad \wedge
\qquad T_c>\left(\frac{T_{GR}^{sph}\zeta}{T_{CR}}\right)^{\frac{q}{2q-1}}T_{Cr}\wedge T_c>T_N\ . \label{cS1}
\end{eqnarray}

Note
that the simultaneous satisfaction of the two latter conditions related with the temperature is necessary because for 
\[
q>\ln(T_N/T_{Cr})/\ln[T_N^2/(T_{GR}^{sph}\zeta)]\simeq 0.2\; ,\] 
we obtain 
\[\left(T_{GR}^{sph}\zeta/T_{CR}\right)^{q/(2q-1)}T_{Cr}<T_N\; .
\]

The intersection of conditions can be written as two conditions for two different ranges of $q$. However, numerically, the expression presented in Equation~\eqref{cS1} is easier to compute. 

To conclude this brief analysis, it is important to understand that the condition \eqref{cS1} is a necessary condition for a viable electroweak baryogenesis scenario to take place during the first STT cosmological epoch.
Even without further enquiries, it becomes clear that a modification of the  underlying gravitational theory 
implies a change in the time--temperature relation  of the evolving cosmological model, altering the conditions that govern the interplay  between the rates of the interactions generating baryon asymmetry, and the expansion rate of the Universe. Therefore the equilibrium of the former does not exactly occur as in the GR standard model, and there are consequences for the baryogenesis mechanisms that have been devised, as we will further see in what follows.

\subsection{Scalar--Tensor GUT Baryogenesis}

The previous analysis of GUT baryognesis implicitly relied on specific assumptions; in particular, it was assumed that early inflation cannot occur after the period during which the baryon asymmetry develops. As previously  mentioned, such a period of accelerated expansion would erase any asymmetry that might have been previously produced. Furthermore, it is important to emphasize that the mechanism devised may be seen as a first approximation to more realistic GUTs~\cite{Kolb:1990vq}. 

Bearing this in mind, the Boltzmann equations that rule the evolution of the baryon asymmetry of the model presented are given by
\begin{align}
\dot{X}&=-\Gamma_D\,(X-X_{EQ}),\label{BoltzmannX1}\\
\dot{B}&=\varepsilon\Gamma_D\,(X-X_{EQ})-(\Gamma_{ID}+2\Gamma_S)\,B\,,\label{BoltzmannX2}
\end{align}
where
\begin{equation}
X=8\,g_{\ast a}N_X\\
X=88\,g_{\ast a}N_{X_{EQ}}\\
B=8\,g_{\ast a}\epsilon N_B \; , \label{rescaled B}
\end{equation}
$g_{\ast a}$ represents the radiation degrees of freedom during the epoch of interest, $\epsilon$ parametrizes the degree of CP violation, $N_X$ is the number of a generic self-conjugate supermassive boson, and $N_{X_{EQ}}$ is its corresponding equilibrium number. Moreover, 
\begin{equation}
N_B=\frac{N_{b}-N_{\bar{b}}}{2} \ ,
\end{equation}
is the mean net baryon number, and $\Gamma_j$, with $j=D,~ID, S$, is the decay rate of the supermassive boson $X$, its inverse decay rate and the baryon-non-conserving (BNC) scattering rate, respectively. 

Now, in order to proceed, it is useful to introduce a new dimensionless, dynamical variable defined by $\bar z=m_X/T$, where $m_X$ is the mass of the generic supermassive boson $X$. In terms of this variable, the rescaled equilibrium number of supermassive bosons $X$ is given by
\begin{equation}
X_{EQ}=2a(\bar z)\,,\label{xeq}
\end{equation}
where
\begin{eqnarray}
a(\bar z)\equiv\frac{\bar z^2}{2}K_2(\bar z)\simeq\Biggl\{
\begin{array}{ll}1&\textrm{if}~\bar z\ll1\\
\frac{1}{2}\sqrt{\frac{\pi}{2}}\bar z^{3/2}\exp(-\bar z)&\textrm{if}~\bar z\gg1
\end{array} \ , 
\end{eqnarray}
and $K_2(\bar z)$ is the modified Bessel function of the second kind. In turn, the thermal expansion rate of the Universe for the first and second epochs can be written respectively as
\begin{eqnarray} 
\Gamma_{e_i}(\bar z)=\Biggl\{\begin{array}{ll}\Gamma_{e_1}(T)=\frac{\sqrt{g_{\ast}}}{0.602 m_{Pl_0}}\frac{m_X^2}{\zeta}\Big(\frac{\bar z_c}{\bar z}\Big)^{1/q-2}\bar z^{-2}\,,&\textrm{if}~0\leq \bar z\leq \bar z_c\\
\Gamma_{e_2}(T)=\frac{\sqrt{g_{\ast}}}{0.602m_{Pl_0}}\frac{m_X^2}{\zeta}\bar z^{-2}\, ,&\textrm{if}~\bar z_c\leq \bar z\leq \bar z_{eq}
\end{array} \ ,
\label{expansionrategutfinal1}
\end{eqnarray}
where $i=1,~2$ stand for the first and second radiation-dominated epochs, respectively, and $z_{c}=T_c/m_X$, $z_{eq}=T_{eq}/m_X$ are, respectively, the $\bar z$ assigned to the time at which the transition for the second epoch occurs and the time of matter--radiation equality. Furthermore, it is possible to establish the following relation
\begin{equation}
\textrm{d}t=\frac{\textrm{d}\bar z}{\bar z\Gamma_{e_i}(\bar z)}\label{diferentialgut}\ .
\end{equation}
     
Analogously to what was done for GR, it is also useful to define a quantity 
\begin{equation}
K_{STT_i}\equiv\left(\frac{\Gamma_D}{\Gamma_{e_i}}\right)_{\bar z=1}=\Biggl\{\begin{array}{ll}K_{STT_1}=K_{GR}\,\zeta \,\bar z_c^{\frac{2q-1}{q}}&\textrm{if}~0\leq \bar z\leq \bar z_c\\
K_{STT_2}=K_{GR}\,\zeta&\textrm{if}~\bar z_c\leq \bar z\leq \bar z_{eq} 
\end{array}
\,,\label{Kgutfinal}
\end{equation}
which measures the effectiveness of the decays at the crucial epoch $\bar z=1$ ($m_X=T$), when the self-conjugated supermassive bosons $X$ are forced to diminish in number if they are to stay in equilibrium. It can be seen that the substitution of the definition of  Equation~\eqref{Kgutfinal} in the differential relation of Equation~\eqref{diferentialgut} gives
\begin{equation}  
\textrm{d}t=\frac{K_{STT_i}}{\Gamma_D(\bar z=1)}\, f_i(\bar z)\,\textrm{d}\bar z\,,\label{differentialGUT2}
\end{equation}
where
\begin{equation}
f_i(\bar z)\equiv \bar z^{-1}\frac{\Gamma_{e_i}(\bar z=1)}{\Gamma_{e_i}(\bar z)}=\Biggl\{\begin{array}{ll}f_1(\bar z)=\bar z^{\frac{1-q}{q}}&\textrm{if}~0\leq \bar z\leq \bar z_c\\
f_2(\bar z)=\bar z&\textrm{if}~\bar z_c\leq \bar z\leq \bar z_{eq}
\end{array} \ .
\end{equation}

The subsequent substitution of the differential relation \eqref{differentialGUT2} and of the rescaled equilibrium number of bosons $X$ \eqref{xeq} into the set of Boltzmann Equations \eqref{BoltzmannX1}--\eqref{BoltzmannX2}, which amounts to a change of variable, provides
\begin{align}
X'&=-K_{STT_i}f_i\,[\gamma_D\,(X-2a)+\gamma_A\,(X^2-4a^2)]\, , \label{BoltzmannX11}\\
B'&=K_{STT_i}f_i\,[\varepsilon\gamma_D\,(X-2a)-(\gamma_{ID}+\gamma_S)\,B]\, , \label{BoltzmannX22}
\end{align}
where the prime $'$ stands for differentiation with respect to $\bar z$ and $\gamma_j\equiv \Gamma_j(\bar z)/\Gamma_D(\bar z=1)$ such that 
\begin{align}
\gamma_D&=\frac{\sqrt{2}}{\sqrt{\bar z^{-2}+1}}\simeq\Biggl\{\begin{array}{ll}\bar z&\textrm{if} \ \bar z\lesssim1\\
1&\textrm{if} \ \bar z\gtrsim1\end{array} \ ,\label{gammaD}\\
\gamma_{ID}&
=2\frac{a(\bar z)}{\sqrt{\bar z^{-1}+1}}\simeq\Biggl\{\begin{array}{ll}2z&\textrm{if} \ \bar z\lesssim1\\
\sqrt{\frac{\pi}{2}}\bar z^{3/2}\exp(-\bar z)&\textrm{if} \ \bar z\gtrsim1 \label{gammaID}
\end{array}\ , \\
\gamma_S&=\frac{A\alpha_X}{\bar z(1+\bar z^2)^2}\simeq\Biggl\{\begin{array}{ll}A\alpha_X \bar z^{-1} &\textrm{if} \, \bar z\lesssim1\\
\frac{A\alpha_X}{\bar z^5}&\textrm{if}\, \bar z\gtrsim1 \label{gammaS}
\end{array} \ ,\\
\gamma_A&\simeq\Biggl\{\begin{array}{ll}\frac{\alpha_X^2}{\alpha_H}\bar z^{-1} &\textrm{if}\ \bar z\lesssim1\\
\frac{\alpha_X^2}{\alpha_H}\bar z^{-3} &\textrm{if} \ \bar z\gtrsim1 \label{gammaA}\end{array} \ ,
\end{align}
where $A$ is a numerical factor, which is introduced in order to mimic the characteristics of more realistic GUTs, such as the number of scattering channels. Regarding these definitions, it is possible to see that the freezing relation $\Gamma_{e_i}(\bar z_{f_{ji}})\simeq\Gamma_{j}(\bar z_{f_{ji}})$, where $\bar z_{f_{ji}}$ is the value of 
$\bar z$ at which the $j$-process freezes during the epoch $i$, can be cast in the simple form 
\begin{equation}
f_i(\bar z_{f_{ji}})=K_{STT_i}\gamma_j(\bar z_{f_{ji}})\label{fiz}\,.
\end{equation}

It becomes apparent from Equation \eqref{Kgutfinal} that  the parameter $K_{STT_i}$ conveys the impact of the modification on the expansion rate, and consequently on the time--temperature relation, of the STT toy model under consideration. This is reflected in the system of 
Boltzmann Equation \eqref {BoltzmannX22}. The differences with respect to GR that  alter the generation of baryon asymmetry in STT clearly arise during the first of the three epochs, when the JBD dilatonic scalar field dominates. A comprehensive and more detailed exploration of this subject  will be expounded  in a forthcoming work.

\section{Conclusions}\label{sec:conclusion}

The generation of the observed baryon asymmetry must have been accomplished during the first stages of the primordial Universe to be compatible with the large-scale homogeneity and isotropy of the evolving Universe. Since shortly after the Planck epoch the physics is well described by the standard model of cosmology and the standard model of particles, one cannot reconstruct a reasonable mechanism for baryogenesis without taking into account these descriptions of the Universe. A viable theoretical explanation for the origin of the baryon asymmetry must always ensure that the Sakharov criteria are fulfilled. In this brief review of the subject, we revisited two well-known mechanisms that aim to solve the matter--antimatter asymmetry problem, namely, the electroweak baryogenesis (EWB) and baryogenesis achieved via GUT. Although both proposals are built around the second Sakharov criteria, i.e., CP violation, from the gravitational viewpoint, one has to focus on the third Sakharov criterion that concerns departure from equilibrium. 
To both explore and illustrate these issues, we considered a different theory of gravity, namely, scalar--tensor theory, and presented a model for the cosmological study of the baryon generating mechanism within this extended framework. 

The two mechanisms that were addressed display an intrinsic interplay between the expansion rate of the Universe and some of the core quantities that govern how these mechanisms lead to breaking in the matter--antimatter symmetry. 
Regarding the EWB, we showed that there is a high relation between the expansion rate of the Universe and the temperature. This connection was then explored in the STT description, where, through a simple analysis, we concluded that the sphaleron bound could be weakened,
thus allowing a more effective EWB.  
In what concerns the GUT baryogenesis, the departure from equilibrium was also shown to play a crucial role, in particular, within the STT framework, in which we briefly explored a toy model that clearly renders that the early stage during which the Brans--Dicke-like scalar field dominates the expansion plays a significant role and introduces differences with regard to the corresponding GR mechanism. This is representative of the types of modifications of the baryogenesis processes that are to be found when considering extended theories of gravity. 
\vspace{6pt}

\authorcontributions{
Formal analysis, D.S.P., J.F., F.S.N.L. and J.P.M.; Investigation, D.S.P., J.F., F.S.N.L. and J.P.M.; Writing—original draft, D.S.P., J.F., F.S.N.L. and J.P.M. All the authors have substantially contributed to the present work. All authors have read and agreed to the published version of the manuscript.}

\funding{This research was funded by the Funda\c{c}\~{a}o para a Ci\^{e}ncia e a Tecnologia (FCT) from the research grants UIDB/04434/2020, UIDP/04434/2020 and PTDC/FIS-AST/0054/2021.}

\institutionalreview{Not applicable.}

\informedconsent{Not applicable.}

\dataavailability{Not applicable. }



\acknowledgments{D.S.P., F.S.N.L. and J.P.M. acknowledge funding from the research grants UIDB/04434/2020, UIDP/04434/2020 and PTDC/FIS-AST/0054/2021. JPM further thanks the funding of the research grant EXPL/FIS-AST/1368/2021.
F.S.N.L. acknowledges support from the Funda\c{c}\~{a}o para a Ci\^{e}ncia e a Tecnologia (FCT) Scientific Employment Stimulus contract with reference CEECINST/00032/2018. 
 }

\conflictsofinterest{The authors declare no conflicts of interest.} 
\clearpage
\abbreviations{Abbreviations}{
The following abbreviations are used in this manuscript:\\

\noindent 
\begin{tabular}{@{}ll}
EWB & Electroweak Baryogenesis\\
GUT & Grand Unification Theories\\
STT & Scalar--Tensor Theories\\
FLRW & Friedmann--Lema{\i}tre--Robertson--Walker \\
FD & Fermi--Dirac \\
BE & Bose--Einstein \\
JBD & Jordan--Brans--Dicke \\
BAU & Baryon asymmetry of the Universe \\
\end{tabular}
}

\begin{adjustwidth}{-\extralength}{0cm}

\reftitle{References}

\PublishersNote{}
\end{adjustwidth}
\end{document}